\documentclass[final,3p,times]{elsarticle}
\usepackage{amssymb}
\usepackage{graphics}
\usepackage{dcolumn}
\usepackage{bm}
\usepackage{array}
\usepackage{booktabs}
\usepackage{amsmath}
\usepackage{multirow}
\usepackage{esint}
\usepackage{color}
\usepackage{graphicx}
\usepackage{epstopdf}
\usepackage{verbatim}
\usepackage{float}
\usepackage{fancyvrb}
\usepackage[export]{adjustbox}
\usepackage{mathrsfs}  
\usepackage[utf8x]{inputenc}
\usepackage[vietnamese=nohyphenation]{hyphsubst}
\usepackage[english,vietnamese]{babel}
\usepackage[vietnamese]{babel}
 
\usepackage{gensymb}
\usepackage{multicol}
\DeclareMathOperator*{\argmax}{arg\,max}

\usepackage{amsthm}
\newtheorem{remark}{Remark}
\usepackage{soul}
\usepackage{caption}
\usepackage{subcaption}
\usepackage{scalerel}
\usepackage{tikz}
\usetikzlibrary{svg.path}
\definecolor{orcidlogocol}{HTML}{A6CE39}
\tikzset{
  orcidlogo/.pic={
    \fill[orcidlogocol] svg{M256,128c0,70.7-57.3,128-128,128C57.3,256,0,198.7,0,128C0,57.3,57.3,0,128,0C198.7,0,256,57.3,256,128z};
    \fill[white] svg{M86.3,186.2H70.9V79.1h15.4v48.4V186.2z}
                 svg{M108.9,79.1h41.6c39.6,0,57,28.3,57,53.6c0,27.5-21.5,53.6-56.8,53.6h-41.8V79.1z M124.3,172.4h24.5c34.9,0,42.9-26.5,42.9-39.7c0-21.5-13.7-39.7-43.7-39.7h-23.7V172.4z}
                 svg{M88.7,56.8c0,5.5-4.5,10.1-10.1,10.1c-5.6,0-10.1-4.6-10.1-10.1c0-5.6,4.5-10.1,10.1-10.1C84.2,46.7,88.7,51.3,88.7,56.8z};
  }
}
\newcommand\orcidicon[1]{\href{https://orcid.org/#1}{\mbox{\scalerel*{
\begin{tikzpicture}[yscale=-1,transform shape]
\pic{orcidlogo};
\end{tikzpicture}
}{|}}}} 
\usepackage{hyperref}
\hypersetup{
  pdftitle={xxxx},
  pdfauthor={xxxx}
}
\journal{Computer Methods in Applied Mechanics and Engineering}
\begin{document}
\begin{frontmatter}
\author[1]{Jonathan Oleson \orcidicon{0009-0008-6196-8193} }
\author[1]{Abhishek Keripale \orcidicon{0009-0009-2127-6536} }
\author[1]{Shashank Pathrudkar \orcidicon{0000-0001-8546-8056} }
\author[1,5]{Susanta Ghosh  \orcidicon{0000-0002-6262-4121} \corref{cor1}}
\ead{susantag@mtu.edu}
\cortext[cor1]{Corresponding author}
\affiliation[1]{organization={Department of Mechanical and Aerospace Engineering},
addressline={Michigan Technological University},
postcode={MI},
country={USA}}

\affiliation[5]{organization={The Center for Artificial Intelligence at the Institute of Computing and Cybersystems},
addressline={Michigan Technological University},
postcode={MI},
country={USA}}
\title{Accelerating Atomistic-Continuum Simulations of Multilayer 2D Materials with Functional Principal Component Analysis-Based Machine Learning}
\selectlanguage{english}
\begin{abstract}
Atomistic-continuum simulations of complex wrinkling deformations in large-area multilayer two-dimensional (2D) materials are computationally expensive. To accelerate these simulations, this work presents a machine-learning (ML) framework that first constructs a low-dimensional representation of the deformation field, then learns the deformation in this reduced space. Dimensionality reduction of deformed 2D membranes is challenging because of the extremely high dimensionality of discretized deformation fields, the need to preserve surface continuity and prescribed geometric constraints, the stochastic nature of wrinkling patterns, and the limited availability of training data. We address these challenges using two-dimensional functional principal component analysis (FPCA). The proposed FPCA representation reduces deformation fields containing millions of degrees of freedom to latent spaces with only tens of dimensions while preserving the continuity of the deformed surface. In addition, prescribed homogeneous boundary constraints are satisfied exactly through an appropriate choice of basis functions in the FPCA representation. A neural network is then trained to predict the principal components from the system and loading parameters. The resulting ML-predicted deformation fields are used as high-quality initial configurations for the atomistic-continuum solver, which subsequently performs energy minimization to obtain equilibrium solutions. The framework is demonstrated for multilayer graphene subjected to uniaxial compression, biaxial compression, and combined biaxial compression and simple shear. For the most computationally intensive large-strain cases, the proposed approach achieves speedups of at least \(6.8\times\) while converging to final energies comparable to, and in some cases lower than, those obtained using conventional atomistic-continuum initialization. Given the large system sizes considered, ranging from millions to several billions of atoms, the resulting computational acceleration is substantial.  The proposed framework provides a general strategy for accelerating computationally intensive physics-based simulations whose high-dimensional solution fields exhibit low-dimensional structural patterns.
\end{abstract}
\end{frontmatter}

\selectlanguage{english}
\section{INTRODUCTION}\label{sec:Introduction}
Two-dimensional (2D) materials have gained enormous interest among researchers since the breakthrough of graphene due to their promising physical and electrical properties \cite{novoselov2004electric}. The family of 2D materials has grown with the advent of materials like transition metal dichalcogenides, hexagonal boron-nitride, MXenes, etc., and now spans a wide range of compositions from the periodic table  \cite{mas20112d,gupta2015recent,geng2018recent, zhang2015ultrathin, gogotsi2019rise}. 2D materials consist of a two-dimensional sheet-like arrangement of constituent atoms forming a strong lattice, flexible in bending yet very stiff in-plane. Multi-layered nanotubes and 2D materials have also shown potential benefits and applications similar to their single-layered forms \cite{poncharal1999electrostatic,shahil2012thermal,li2017layer,yankowitz2019tuning}. The material properties of these nanomaterials are strain-tunable; applications of mechanical strains to these materials that result in complex deformations (such as twisting, rippling, and wrinkling) strongly affect their properties, including their electronic and optical properties \cite{falvo1997bending,  williams2002torsional}. In particular, torsion and bending in nanotubes and compression in layered 2D materials have been widely studied owing to the useful electromechanical response induced by the resulting deformations \cite{yankowitz2019tuning,williams2002torsional,papadakis2004resonant,de2013carbon,pereira2009strain}. 

Conventional atomistic simulation methods, such as first-principles calculations or molecular dynamics simulations (see \cite{Wei2009PRB,de2012density,zhao2009size} for graphene), can be used to reliably simulate small-scale samples of 2D materials but are computationally prohibitive for large single-layered and multi-layered samples containing up to a billion atoms \cite{Zang2013NatMat}. Given the huge computational requirements of all-atom models for these materials, and given their thin shell-like behavior \cite{Yakobson1996PRL}, these systems are also modeled in terms of only continuum mechanics \cite{Pantano_electro, Pantano2004JMPS}, which ignores the underlying atomistic physics. To leverage the computational advantages of continuum models at a large length scale while respecting atomistic physics, continuum constitutive models were developed based on interatomic potentials and using the Cauchy-Born rule. We refer to them here as atomistic-continuum models. For example, in \cite{Arroyo2002}, the authors presented a membrane model for carbon nanotubes, which we refer to as the \emph{surface} model. The surface model has developed a continuum elastic potential from the interatomic potential of graphene by exploiting the extended Cauchy-Born rule. This model was successfully validated against all-atom simulations and provided a means to simulate multi-million-atom graphene-based systems at two orders of magnitude lower computational cost versus purely atomistic simulations \cite{Arroyo2002, Arroyo2004IJNME}. While this model dramatically reduces the number of degrees of freedom when discretized with finite elements, the number of quadrature points necessary to resolve the length scale of van der Waals interactions is not significantly less than the number of atoms. This limits the computational efficiency of the model, though it remains faster than all-atom models. To overcome this challenge, a three-dimensional model was presented in \cite{GhoshJMPS2013}, called the \emph{foliation} model, where continuum constitutive models were developed from interatomic interactions for both the bonded and the non-bonded interactions. The foliation model's elastic energy formulation for the non-bonded (van der Waals) interactions between the layers is motivated by the observation that the spacing between neighboring layers of multilayer graphene and multiwalled carbon nanotubes (MWCNTs) varies smoothly. For the bonded interactions, the foliation model uses the surface model's elastic energy formulation for each layer of multilayer graphene or MWCNTs. The leaves of foliation represent individual graphene layers of the multilayer systems. The stretch and bending kinematics of the leaf can be related to the deformation of covalent bonds within a layer, which constitute the continuum energy associated with the bonded interactions. The foliation model measures the local density of graphene sheets transverse to the leaves/layers and uses it to evaluate the van der Waals interaction energy. We elaborate on the kinematics of the foliation model in Section \ref{sec:background_foliation}. We use the foliation model to simulate training data for our machine learning model. The foliation model is relatively computationally efficient, approximately ten times faster than the surface model. Though these atomistic-continuum models are orders of magnitude faster than the all-atom models, they still require considerable amounts of high-performance computing resources for large-scale simulations, hindering the exploration of the mechanics of two-dimensional materials. In this work, we employ ML algorithms to accelerate the state-of-the-art atomistic-continuum (foliation) model to predict the deformation of two-dimensional materials.

Various ML models have been developed to predict low-dimensional properties of materials. Some ML models specifically for carbon nanotubes, such as \cite{vcanadjija2021deep, vivanco2022machine, kovsmerl2022predicting}, predict their stress-strain behavior and physical properties. However, our objective is to predict the wrinkling deformation of multilayer 2D materials that, in a discretized form, lies in a very high-dimensional space. For instance, multilayer graphene or multi-walled carbon nanotubes require millions of degrees of freedom to describe their deformation \cite{GhoshJMPS2013,Arias2008PRL}. There are various ML models (especially encoder-decoder-type frameworks) that successfully predict high-dimensional quantities \cite{hanakata2020forward,balu2019deep}; however, these deep-learning models also require high-dimensional input data for training. The present work seeks to predict high-dimensional deformation data (about a fraction of a million-dimensional) from a few scalar inputs that define the layer geometry of the material and the loading parameters. Currently, no deep learning model can accurately predict such high-dimensional deformations from a few low-dimensional system inputs while training on a small dataset. We address this issue by first reducing the dimension of the deformed configuration. The deformed configuration of a 2D layer can be seen as a 2D nonlinear manifold. Various nonlinear dimensionality-reduction techniques enable the representation of nonlinear manifolds in low-dimensional space (called the \textit{latent space}). These techniques include Isomap, t-distributed stochastic neighbor embedding, locally linear embedding, UMAP, Autoencoders, etc \cite{Book_NonlinearDR_Lee_Verleysen}. However, it is often infeasible to interpolate in the latent space or to recreate the high-dimensional data from the low-dimensional representation after interpolation in the latent space. The solution must also satisfy the prescribed displacement boundary conditions exactly, a requirement that is not straightforward to enforce in conventional deep-learning models. Furthermore, it is crucial for the high-dimensional deformed configuration reconstructed from the latent space to maintain the smoothness of a continuum solution. In view of these requirements, we adopt functional principal component analysis (FPCA) to reduce the dimensionality of the deformation data. FPCA provides a functional representation of data \cite{yao2005, ramsay2007applied}. This approach is analogous to the KL expansion, representing a stochastic process as a linear combination of infinite orthogonal basis functions \cite{karhunen1947ueber, wang2008karhunen}. To reduce the dimensionality, the deformation is approximately represented via a limited number of terms of this expansion. In FPCA, the deformation fields are represented using a set of two-dimensional basis functions chosen to satisfy the prescribed displacement boundary conditions exactly. Consequently, any truncated FPCA expansion constructed from these basis functions also satisfies the boundary conditions exactly. The proposed machine learning model consists of two main steps: first, perform dimensionality reduction on the data through functional principal component analysis; second, use a neural network to map from system inputs to the reduced-dimensional representation of the deformation. The inputs to the neural network model are the system geometry parameters and the loading parameter (applied strain). We demonstrate the capabilities of our model by predicting the deformed configuration of multilayer graphene under various loading conditions. The wrinkling patterns in simulated multilayer 2D materials are highly sensitive to small random perturbations in the initial configuration because of their very low bending stiffness. Consequently, repeated simulations of the same system can converge to different wrinkle configurations while preserving the same overall morphological characteristics. Although these realizations may differ locally in the spatial arrangement of wrinkles, they typically correspond to energetically comparable equilibrium states. Consequently, the ML-predicted deformation for an unseen system is unlikely to reproduce a specific equilibrium configuration exactly. Moreover, the energy associated with the deformation directly predicted by ML may differ substantially from that of the corresponding equilibrium state. Therefore, we use the ML-predicted deformations as informed initial guesses for the atomistic-continuum solver, which subsequently performs energy minimization to obtain an equilibrium solution. The resulting hybrid framework, combining ML inference with energy minimization via the foliation model, is hereafter referred to as the ML-foliation model. The accuracy and computational efficiency of the proposed ML-foliation model are evaluated extensively. 

The main contributions of this work include: (i) a functional dimensionality-reduction formulation that applies FPCA directly to two-dimensional deformation manifolds, thereby overcoming the limitations of FPCA applied to one-dimensional curves while exactly enforcing the prescribed boundary constraints; and (ii) a hybrid ML-physics approach, referred to as the ML-foliation model, in which ML-predicted deformation fields serve as initial guesses for the foliation model, enabling rapid convergence to an energy-minimized atomistic-continuum solution. 

The remainder of the paper is organized as follows. In Section \ref{sec:background_foliation}, we explain the kinematics of the foliation model, which is the atomistic-continuum model used to generate the training data for our machine learning model and minimize the systems obtained from the ML predictions. In Section \ref{sec:DML_model}, we provide an overview of the proposed ML model, specifically details of the proposed dimensionality-reduction technique and the deep neural network for learning in the reduced-dimension space. Section \ref{sec:dat_gen_aug} describes data generation and augmentation for the ML model. In Section \ref{sec:results}, we validate the ML model by comparing the ML-obtained deformed configuration after minimization against the atomistic-continuum model in terms of energy and deformation pattern morphology. We also quantify the reduction in computational cost by using the ML framework. The next section includes comments on the interpretability and explainability of the proposed model. We conclude with a summary and discussion of the contributions of this work and applicability to other areas of research.

\section{Background: The foliation model}\label{sec:background_foliation}

In this section, we describe the atomistic-continuum model used to simulate multilayer graphene for the proposed manifold learning model. The theory stems from the idea that the crystalline membranes corresponding to the layers of multilayer graphene (or walls of MWCNTs) readily present themselves as parallel two-dimensional (2D) sub-manifolds embedded in three-dimensional (3D) space, as discussed in \cite{GhoshJMPS2013} by Ghosh and Arroyo. The collection of these sub-manifolds is referred to as a \textit{foliation} in differential geometry, with the sub-manifolds themselves referred to as leaves \cite{Cohen1974AMM} \cite{Rovenski1995} \cite{Candell1999}. For further details on the foliation model, one may refer to the paper by Ghosh and Arroyo.

\begin{figure}[htbp]
    \centering
    \includegraphics[width=0.75\linewidth]{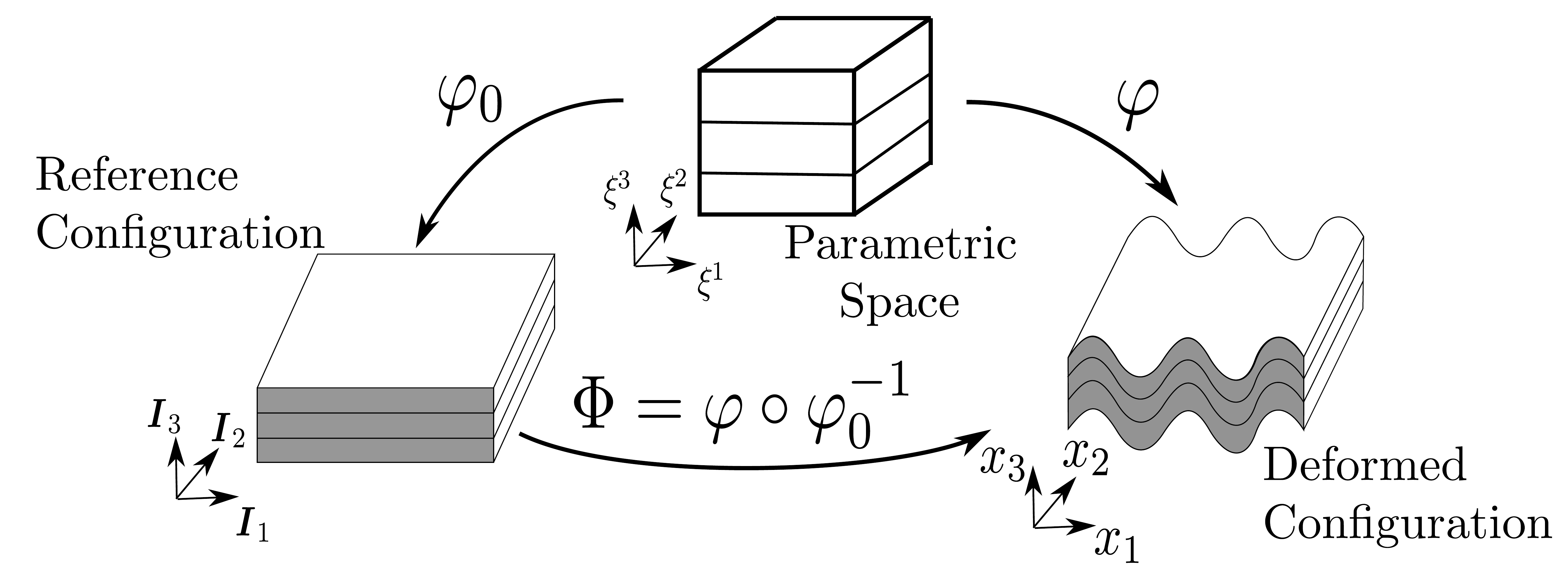}
    \vspace{-0pt}
    \caption{Kinematics: the configuration spaces, deformation maps, and coordinates for multilayer graphene.}
    \vspace{-10pt}
    \label{fig:fol}
\end{figure}

The overall deformation map $\Phi$ is a composition of the injective map from the parametric space to the reference configuration (denoted as $\varphi_0$) and the map from the parametric space to the Euclidean space $\mathbb{R}^3$ in which the deformed configuration resides (denoted as $\varphi$). This deformation map is defined as $\Phi = \varphi \circ \varphi_0^{-1}$ (a schematic of all configurations and the corresponding maps is shown in Figure \ref{fig:fol}). 
For the defined deformation map, $\Phi$, the deformation gradient can be expressed as 
\begin{equation}
\textbf{\textit{F}} = D_{\xi}\varphi \, D_{\xi}\varphi_0^{-1}
\end{equation}
This deformation gradient can be used to obtain the Cauchy-Green strain tensor of the leaves of the foliation as 
\begin{equation}
\textbf{\textit{C}}^l = (D_{\xi}\varphi_0^l)^{-T}\textbf{\textit{g}}\,(D_{\xi}\varphi_0^l)^{-1}
\end{equation}
where the $l$ superscript denotes a parameter for a given leaf, and $\textbf{\textit{g}}$ is the metric tensor of each leaf at each point, with its components expressed as $\textit{g}_{\alpha\beta} = \left\langle\frac{\partial\varphi^l}{\partial\xi^{\alpha}},\frac{\partial\varphi^l}{\partial\xi^{\beta}}\right\rangle$, where $\alpha,\beta = 1,2$ and $\langle \cdot,\cdot \rangle$ denotes the Euclidean scalar product in $\mathbb{R}^3$. Similarly, the curvature tensor is defined as the pullback of the Weingarten map and is expressed as
\begin{equation}
 \mathscr{K}= (D_{\xi}\varphi_0^l)^{-T}\textbf{\textit{k}}\,(D_{\xi}\varphi_0^l)^{-1}
\end{equation}
with $\textbf{\textit{k}}$ being the curvature of each leaf, obtained as components of the Weingarten map and expressed as $\textit{k}_{\alpha \beta} = \textbf{\textit{n}},\textbf{\textit{g}}_{\beta,\alpha}$, where $\textbf{\textit{n}}$ is the unit normal vector at that point, $\alpha,\beta=1,2$, and the comma in the subscript indicates the partial derivative with respect to $\xi^{\alpha}$. 

The energy of the graphene system is composed of the bonded interactions and non-bonded interactions. The discrete bonded interactions are represented by a modified Brenner potential written in terms of atomic deformation measures such as bond lengths and bond angles. For a single graphene layer, a continuum strain energy density, written in terms of the continuum strain measures of the equivalent continuum surface model, can be derived from discrete bonded interactions using the exponential Cauchy-Born rule \cite{Arroyo2004IJNME}. The continuum strain energy per unit reference surface area of a single layer is represented in terms of the right Cauchy-Green strain and the curvature tensor of the leaves of the foliation as $W(\textbf{\textit{C}}^l,\mathscr{K})$. Note that in the foliated bulk continuum, the surface-bonded energy density $W(\textbf{\textit{C}}^l,\mathscr{K})$ can be evaluated at any point in the body, which is represented as a continuous distribution of leaves. To represent 3D bulk multilayer graphene, the foliation model assigns the thickness $t_0$ to each 2D layer, which corresponds to the equilibrium spacing between graphene sheets. By distributing the bonded energy density per unit reference area over the thickness $t_0$, the volumetric bonded energy density is obtained as $\mathscr{W}(\textbf{\textit{C}}^l,\mathscr{K}) = \frac{1}{t_0}\,W(\textbf{\textit{C}}^l,\mathscr{K})$. 

The inter-layer van der Waals (non-bonded) interaction was modeled using a potential that is a function of the inter-layer distance, $t$. Specifically, a 4-10 LJ potential was obtained by taking the sum of pairwise 6-12 LJ potentials over all atom pairs in the system. The inter-layer potential can be expressed as 
\begin{equation}
    V_{vdW}(t) = -\frac{|V_g(t_0)|}{0.6}\bigg[\bigg(\frac{t_0}{t}\bigg)^4 - 0.4\bigg(\frac{t_0}{t}\bigg)^{10}\bigg]
\end{equation}
where $V_g(t_0)$ represents its value at $t=t_0$, and $t_0$ represents the equilibrium spacing. The effective volumetric van der Waals potential can be expressed as 
\begin{equation} 
    \mathscr{V}(\lambda^t,\textbf{\textit{C}}^l) = \left(\frac{N_w-1}{N_w}\right) \frac{1}{t_0 J^l}\,V_{vdw}(\lambda^t t_0)
\end{equation}
where $N_w$ is the number of layers in the graphene sample, $\lambda^t$ is the transversal stretch representing the relative spacing between the layers with respect to the equilibrium spacing, and $J^l$ represents the Jacobian determinant of the leaf. A complete derivation of the effective volumetric potential for the van der Waals interaction can be found in \cite{GhoshJMPS2013}. Combining the bonded energy and van der Waals interactions, the total internal energy of the system becomes 
\begin{equation}
    \Pi_{int}(\varphi) = \int_{\bar{V}} \left(\mathscr{W}\left(\textbf{\textit{C}}^l,\mathscr{K}\right) + \mathscr{V}\left(\lambda^t,\textbf{\textit{C}}^l\right)\right)\, dV_0
\end{equation}
In addition, given the potentials associated with any applied external forces in the reference domain, denoted as $\Pi_{ext}(\varphi)$, the total potential energy can be expressed as
\begin{equation}
    \Pi_{total}(\varphi) = \Pi_{int}(\varphi) - \Pi_{ext}(\varphi)
\end{equation}
Given the boundary conditions and the external loading, the resulting deformed shape is obtained by minimizing the total energy with respect to the deformation map. A direct Ritz-Galerkin formulation is implemented, where the configuration is approximated using B-splines. Additional details for the numerical implementation and validation of the foliation model are provided in \cite{GhoshJMPS2013}.

The foliation model is used to generate data for the proposed ML model as well as to minimize the energy of the deformed configuration predicted by the ML model.

\section{Proposed Machine Learning Model}\label{sec:DML_model}
In this section, we describe the proposed ML model. The objective of this ML model is to learn the deformation map $\Phi$ for a given boundary condition on a system. The ML model consists of two parts: (i) functional dimension reduction and (ii) supervised learning in the reduced dimension, which are described in sections \ref{sec:dim_red} and \ref{sec:NN}, respectively.  

The undeformed geometry of multilayer graphene is characterized by the thickness of the system defined by the total number of layers, $N_w$, and the location of the graphene layer defined by index $m$. The applied compressive load is denoted by $\Theta$. Together, these three parameters, $N_w$, $m$, and $\Theta$, serve to specify a particular layer within a multilayer graphene system subjected to loading. Accordingly, we let $\mathcal{H}$ denote a map from the space consisting of these geometry and loading parameters to the deformed configuration: $\mathcal{H} : \{ N_w, m, \Theta\} \rightarrow \{\Phi\}$. Note that the outputs of this map $\mathcal{H}$ remain as 2D manifolds even after discretizing the multilayer graphene systems into individual layers. The discretized manifolds are high-dimensional in terms of degrees of freedom, making them difficult to predict directly via standard machine learning models; this challenge in predicting such high-dimensional outputs using ML models is a well-documented phenomenon known as the \emph{curse of dimensionality} \cite{bishop2006pattern}. 

\begin{figure*}[htbp]
    \hspace{0pt}
    \centering
    \includegraphics[width=1\linewidth]{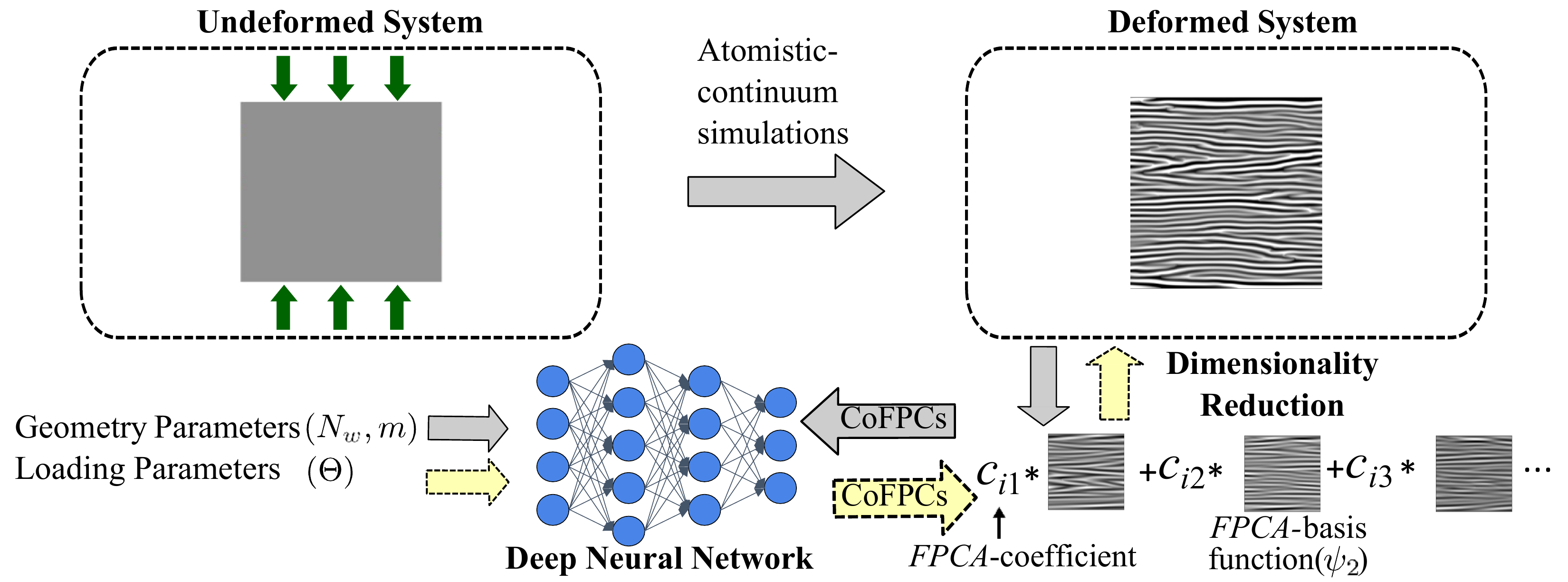}    
    \caption{Schematic of the machine learning framework. Data generation is through atomistic-continuum simulations, dimensionality reduction is through functional principal component analysis, and learning in the reduced dimension is through a deep neural network. Firm arrows show data generation and training, and dashed arrows show inference via the proposed model.}
    \label{fig:DLM_schematic}
\end{figure*}

We address this challenge of high-dimensionality by developing a two-step machine learning model in which we obtain and subsequently learn a low-dimensional representation of the deformed manifold. This two-step machine learning model (outlined in Figure \ref{fig:DLM_schematic}) draws inspiration from our previous works on predicting high-dimensional electronic fields of quasi-one-dimensional materials under mechanical strain \cite{shashank_prb_2022} and high-dimensional deformed surfaces of MWCNTs under torsion and bending \cite{yadav2021interpretable}.

\begin{figure}[htbp]
    \centering 
    \includegraphics[width=0.35\linewidth]{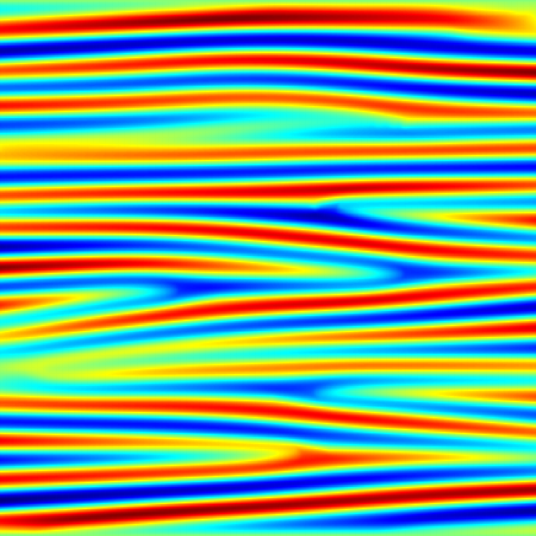}
    \includegraphics[width=0.35\linewidth]{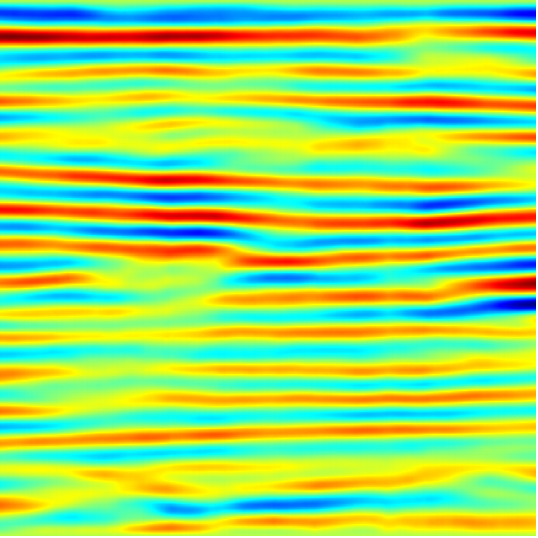}
    \includegraphics[width=0.35\linewidth]{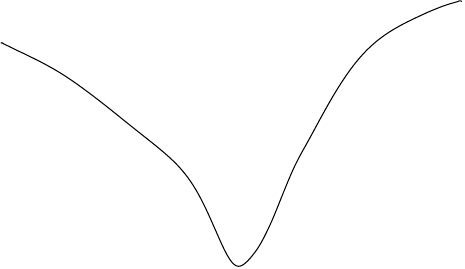}
    \includegraphics[width=0.35\linewidth]{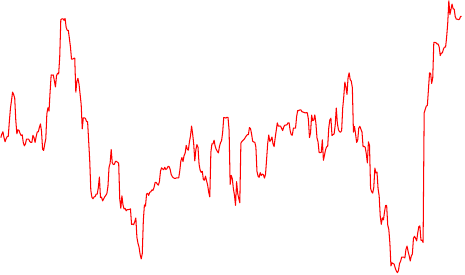}
    \caption{A comparison of the original foliation results (left) and a prediction using a 1D FPCA model (right) for 13-layer graphene under 5\% uniaxial compression. Images in the first row show the out-of-plane deformation at the top layer, and images in the second row show the numerical derivative of out-of-plane deformation with respect to $x_1$ on the $x_1-x_3$ plane cross-section at the center. The derivative of the prediction indicates poor continuity between the aggregated 1D manifolds for the 1D FPCA prediction.}
    \label{fig:discontinuities_gpn}
\end{figure}

The dimensionality-reduction technique proposed here differs from the approach described in \cite{yadav2021interpretable}, which approximates a 2D manifold as a collection of 1D manifolds on which manifold learning is performed. In the 1D approach, the geometry of the MWCNT is represented as a collection of curves rather than the continuous 2D surfaces themselves. This approximation can lead to loss of continuity in the rippling pattern when the manifold is reconstructed by aggregating the curves, limiting the applicability of the model proposed in \cite{yadav2021interpretable}. The problem is exacerbated when applying 1D FPCA to multilayer graphene systems. Examining a sample multilayer graphene prediction using a 1D FPCA approach reveals that the derivative of out-of-plane deformation with respect to the component in the direction of curve aggregation exhibits poor continuity (Figure \ref{fig:discontinuities_gpn}).
In this work, we directly perform functional dimensionality reduction on the 2D manifolds (layers of multilayer graphene). 

Furthermore, MWCNTs are highly constrained systems that produce similar final deformations for various random initializations; however, multilayer graphene is a much less constrained system, and the final deformations depend more heavily on the random initialization. Thus, the machine learning model would function poorly as a \textit{surrogate} model for multilayer graphene. Instead, the model is better-suited to serve as an \textit{accelerator} by learning the pattern from a few sample systems, producing initial guesses for the final deformed configuration of interest, then minimizing the energy of the guess via the atomistic-continuum code (called the foliation model \cite{GhoshJMPS2013}). Subsequently, we assess the performance of the proposed ML-accelerated atomistic-continuum model (referred to as the ML-foliation model) by comparing it against the foliation model, both in terms of the deformation morphology and the energy.

\subsection{Functional dimensionality reduction}\label{sec:dim_red}

Here, we provide an overview of functional principal component analysis (FPCA), the dimensionality-reduction approach employed in the present work. FPCA represents stochastic data via a smooth, functional basis. This basis is theoretically composed of an infinite set of orthogonal functions. The formulation is presented below for the \emph{Kosambi---Karhunen---Loève} expansion (also known as \emph{Karhunen---Loève} expansion or KL expansion)  \cite{karhunen1947ueber, wang2008karhunen, Jorgensen2007, maccone2009deep, Stark1986Book, alexanderian2015brief, tipireddy2020conditional, PHOON2002293}, which is very similar to FPCA.

As mentioned in Sec. \ref{sec:background_foliation}, the parametric surfaces of the ($\xi^1,\xi^2$) plane are analogous to the individual layers in multilayer graphene. Each component of deformation lies in the domain $\Omega_{12} = \mathcal{T}_1\times\mathcal{T}_2$, such that $\Omega_{12} = (\xi^1, \xi^2) \in [0,1]\times[0,1]$. The deformed configuration at the $\xi^3_m$ coordinate is computed on the undeformed surface and given as $\Phi(\xi^1,\xi^2)$. The collection of $N$ such surfaces is represented as ${\{\Phi}_i(\xi^1, \xi^2)\}_{i=1}^{N}$, where $N$ is the number of manifolds obtained from all the simulations used to build the training dataset. Treating $\Phi$ as a stochastic process, it is assumed that $\Phi_i(\xi^1,\xi^2)$ for each layer are sampled from the stochastic process $\Phi(\xi^1,\,\xi^2)$ defined such that its second derivative is square-integrable. This smoothness condition is essential since the energy of graphene membranes is a function of the curvature of a given layer. We suppose that $\Phi(\xi^1,\,\xi^2)$ can take any of the values $\Phi_i(\xi^1,\,\xi^2)\, \in \mathscr{H}^{2}(\mathcal{T}),\; i = 1,\cdots,N$, where $\mathscr{H}^{2}(\mathcal{T})$ is a Hilbert space and $\mathcal{T} = \mathcal{T}_1\times\mathcal{T}_2$. Any geometric constraints, such as periodicity, can be defined as $g(\Phi_i(\xi^1,\,\xi^2))=0$.

We consider the second-order zero-centered processes $\mathcal{X}(\xi^1,\xi^2)$, $\mathcal{Y}(\xi^1,\xi^2)$, and $\mathcal{Z}(\xi^1,\xi^2)$ to represent the three (zero-mean) components of the deformed configuration ${\Phi^1}(\xi^1, \xi^2)$, ${\Phi^2}(\xi^1, \xi^2)$, and ${\Phi^3}(\xi^1, \xi^2)$, respectively. For instance, for the first component of deformation, \[\mathcal{X}(\xi^1,\xi^2) = {\Phi^1}(\xi^1, \xi^2) - \mu^1(\xi^1, \xi^2),\] with $\mathcal{X}_i(\xi^1,\xi^2) = {\Phi^1}_i(\xi^1, \xi^2) - \mu^1(\xi^1, \xi^2), \,  i=1,\dots,N$, where $\mu^1(\xi^1,\,\xi^2)=\frac{1}{n}\,\sum_{i=1}^n\Phi^1_i(\xi^1,\,\xi^2)$ is the mean of ${\Phi^1}_i(\xi^1, \xi^2)$. Similarly, the other two components can be represented through $\mathcal{Y}(\xi^1,\xi^2)$ and $\mathcal{Z}(\xi^1,\xi^2)$. The rest of the analysis is shown only for the first component since it is identical for the others.

We perform dimensionality reduction by first representing the $\mathcal{X}(\xi^1,\xi^2)$ as an infinite linear combination of orthogonal functions by applying the KL expansion as 
\begin{equation} \label{eq:expansion}
    \mathcal{X}(\xi^1,\xi^2)  = \sum_{k=1}^{\infty} \sqrt{\lambda_k} \, \psi_k(\xi^1,\,\xi^2) \, Z_k \; = \sum_{k=1}^{\infty}  \, \psi_k(\xi^1,\,\xi^2) \, \zeta_k
\end{equation}
where $\psi_k(\xi^1,\,\xi^2)$ are deterministic orthogonal functions. $Z_k$ are uncorrelated random coefficients with zero mean and unit variance, and the $\zeta_k$ are uncorrelated RVs with zero mean and variance $\lambda_k$. They are defined as 
\begin{equation} \label{eq:coeff}
    Z_k = \frac{1}{\sqrt{\lambda_k}} \int_{\mathcal{T}_2} \int_{\mathcal{T}_1} \mathcal{X}(\xi^1,\xi^2) \,\psi_k(\xi^1,\xi^2) \,d\xi^1 d\xi^2\, , \quad \zeta_k = \sqrt{\lambda_k}\,Z_k
\end{equation}
The basis functions $\psi_k$ and the uncorrelated random coefficients $Z_k$ together form a bi-orthonormal system \cite{PHOON2002293}. The coefficients $Z_k$ given in equation~(\ref{eq:coeff}) are constructed such that $\mathbb{E}(Z_k)=0$ and $\mathbb{E}(Z_j \, Z_k)=\delta_{jk}$, where $\delta_{jk}$ represents the Kronecker delta. Assuming the covariance function is a continuous, symmetric, non-negative definite kernel and can be diagonalized, then following Mercer's theorem \cite{mercer1909functions}, we have
\begin{equation}
   \int_{\mathcal{T}_2} \int_{\mathcal{T}_1} C((\xi^1, \xi^2),(\vartheta^1,\vartheta^2))\;\psi_k(\xi^1, \xi^2)\,d\xi^1 \,d\xi^2 = \lambda_k\;\psi_k(\vartheta^1,\vartheta^2), \qquad  k = 1, 2, 3, \cdots 
   \label{eq:integral}
\end{equation}
The $\lambda_k$ and $\psi_k$ are the eigenvalues and eigenfunctions
\footnote{
The first eigenfunction $\psi_1$ represents the most dominant mode of variation (denoted by Var) of $\mathcal{X}(\xi^1,\xi^2)$ as \cite{ramsay2007applied}
\begin{equation*}
\psi_{1} = \! \argmax_{\|\psi\|=1} \! \,  \rm{Var} \left( \int_{\mathcal{T}_2} \int_{\mathcal{T}_1} \mathcal{X}\left(\xi^1,\xi^2\right)\psi\left(\xi^1,\xi^2\right)\, d\xi^1 \, d\xi^2 \right), \qquad \text{where}\;  \|\psi\|= \left(\int_{\mathcal{T}_2} \int_{\mathcal{T}_1} \psi\left(\xi^1,\xi^2\right)^2\, d\xi^1 \, d\xi^2\right)^{\frac{1}{2}}
\end{equation*}
The $k$th eigenfunction $\psi_k$ is the dominant mode of variation of $\mathcal{X}(\xi^1,\xi^2)$, orthogonal to $\psi_1,\, \psi_2, \cdots, \psi_{k-1}$, maximizes the variance  as 
\begin{equation*}
\psi_{k} = \! \argmax_{\genfrac{}{}{0pt}{}{\|\psi\|=1,\,\langle \psi, \psi_j \rangle=0}{\text{for} \, j=1,\cdots,k-1}} \!  \rm{Var} \left( \int_{\mathcal{T}_2} \int_{\mathcal{T}_1} \mathcal{X}\left(\xi^1,\xi^2\right)\psi\left(\xi^1,\xi^2\right)\, d\xi^1 \, d\xi^2 \right)
\end{equation*}
where $\langle \psi, \psi_j \rangle \, =\, \int_{\mathcal{T}_2} \int_{\mathcal{T}_1} \psi\left(\xi^1,\xi^2\right)\psi_j\left(\xi^1,\xi^2\right)\, d\xi^1 \, d\xi^2$, for $j = 1,2, \cdots k-1$.
}, 
respectively, of the covariance function $C((\xi^1, \xi^2),(\vartheta^1,\vartheta^2))$ of the random process $\mathcal{X}(\xi^1,\xi^2)$. The eigenvalues $\lambda_k$ are real and arranged in decreasing order, $\lambda_1 \geq \lambda_2 \geq \cdots$. The sum of the squares of the eigenvalues is finite. 

One can solve this Fredholm integral equation of the second kind (\ref{eq:integral}) for the eigenvalues and eigenvectors in a finite-dimensional setting by projecting the data on a finite-dimensional function space. This can be done by numerically solving the Fredholm integral equation or via the singular value decomposition \cite{safta2026kle}. 

To this end, we represent the eigenfunction $\psi(\xi^1,\xi^2)$ in terms of the basis $\mathcal{B}$ of a Hilbert space $\mathscr{H}^{2}(\mathcal{T}_1 \times \mathcal{T}_2)$. A $p$-dimensional basis ($\mathcal{B}$) is chosen following our previous work~\cite{yadav2021interpretable} as
\begin{equation}\label{eq:basis}
\mathcal{B} = \left\{\phi_{l} \in \mathscr{H}^{2}(\mathcal{T}_1 \times \mathcal{T}_2),\, g(\phi_{l})=0,\, l=1,\cdots,p \right\}    
\end{equation}
This basis is selected such that it exactly represents the geometric constraint $g(\phi_l)=0,\, l=1,\cdots,p$. 
In the present experiments, the geometric constraints are the homogeneous boundary conditions of the deformation of multilayer graphene.

\subsubsection{Solution via numerical approximation of Fredholm integral equation}
The $k$th eigenfunction's approximate representation in the basis $\mathcal{B}$ is expressed as
\begin{subequations}\label{eq:eigen_basis_expansion}
\begin{eqnarray}
 \psi_{k}(\xi^1,\xi^2) = \sum_{l=1}^{p} w_l^{(k)} \phi_{l}(\xi^1,\xi^2), \quad w_l^{(k)}\in \mathbb{R}, \quad or \\
 \psi_{k}(\xi^1,\xi^2) = \boldsymbol{\phi}(\xi^1,\xi^2)^{t}\,\mathbf{w}^{(k)} ,\quad \mathbf{w}^{(k)} \in \mathbb{R}^{p}
\end{eqnarray}
\end{subequations}
where $\mathbf{w}^{(k)}$ is a coefficient vector and $\boldsymbol{\phi}(\xi^1,\xi^2)^{t}$ is the vector of Fourier basis functions, both having dimension $p\times1$. Substituting Eq. \ref{eq:eigen_basis_expansion} into Eq. \ref{eq:integral} and exploiting the orthogonality of each basis function leads to the following equation \cite{PHOON2002293}:
\begin{multline}
    \sum_{i=1}^{p}\, w_i^{(k)}\,   \int_{\mathcal{T}_2} \int_{\mathcal{T}_1}
    \int_{\mathcal{T}_2} \int_{\mathcal{T}_1} C((\xi^1, \xi^2),(\vartheta^1,\vartheta^2))\,\phi_i(\xi^1, \xi^2)\,\phi_j(\vartheta^1,\vartheta^2)\,d\xi^1\, d\xi^2\, d\vartheta^1 \,d\vartheta^2 \\ = \lambda_k\,\sum_{i=1}^{p}\, w_i^{(k)}\, \int_{\mathcal{T}_2} \int_{\mathcal{T}_1} \phi_i(\vartheta^1,\vartheta^2)\,\phi_j(\vartheta^1,\vartheta^2)\,d\vartheta^1 \,d\vartheta^2, \qquad  k = 1, 2, 3, \cdots
    \label{eq:integral2}   
\end{multline}
or
\begin{equation*}
    \mathbf{A}\,\mathbf{w}^{(k)} =\, \lambda_k \mathbf{B}\,\mathbf{w}^{(k)} 
\end{equation*}
where the components in the $p\times p$ matrices are 
\begin{eqnarray}
    A_{ji} &=& \int_{\mathcal{T}_2} \int_{\mathcal{T}_1}
    \int_{\mathcal{T}_2} \int_{\mathcal{T}_1} C((\xi^1, \xi^2),(\vartheta^1,\vartheta^2))\,\phi_i(\xi^1, \xi^2)\,\phi_j(\vartheta^1,\vartheta^2)\,d\xi^1\, d\xi^2\, d\vartheta^1 \,d\vartheta^2 \\
    B_{ji} &=& \int_{\mathcal{T}_2} \int_{\mathcal{T}_1} \phi_i(\vartheta^1,\vartheta^2)\,\phi_j(\vartheta^1,\vartheta^2)\,d\vartheta^1 \,d\vartheta^2 
\end{eqnarray}
The above eigenvalue problem is solved using numerical quadrature (as shown in \cite{safta2026kle,PHOON2002293}) by generating samples of the stochastic process at the quadrature points.

\subsubsection{Solution via Singular Value Decomposition}
By projecting the data on the basis $\mathcal{B}$, one can represent the samples of the stochastic process as vectors and obtain an approximate solution for Equation~\ref{eq:integral} via singular value decomposition (SVD) \cite{safta2026kle}, as adopted in the present work.

In this work, we have chosen $\phi_{l}$ as the Fourier Basis and $\mathcal{B}$ as a $p$-dimensional Fourier Basis set that satisfy the geometric constraint exactly. The presently adopted Fourier basis set encodes the geometric constraint of the deformation of multilayer graphene as: $\phi_l(\xi^1 = \{0, 1\},\xi^2=\{0, 1\})=0;\; l = 1, 2, \cdots, p$. The $i$th sample of the stochastic process can be written in the Fourier basis as 
\begin{equation}\label{eq:sample_basis_expansion}
 \mathcal{X}_{i}(\xi^1,\xi^2) = \boldsymbol{\phi}(\xi^1,\xi^2)^{t}\,\mathbf{x}_{i} ,\quad \mathbf{x}_{i} \in \mathbb{R}^{p}
\end{equation}
where $\mathbf{x}_{i}$ is the associated Fourier coefficient vector and $\boldsymbol{\phi}(\xi^1,\xi^2)^{t}$ is the vector of Fourier basis functions having dimension $p\times1$. 

Let the columns of a matrix $D$ be the coefficient vectors $\mathbf{x}_{i}$, thus making $D$ a $p \times N$ matrix. The covariance of $D$ yields the matrix $K$, defined as 
\begin{equation}
    K \, =\, \frac{1}{N-1}\, D\, D^T
\end{equation}
To acquire the eigensystem for $K$, we first take the SVD of the scaled sample matrix $D/\sqrt{N-1}$ as
\begin{equation}
    \frac{1}{\sqrt{N-1}}\,D=U\,\Sigma\,V^T
\end{equation}
where U and V are unitary matrices and $\Sigma$ is a rectangular matrix with non-negative singular values $\sigma_i$ on the diagonal. Rewriting $K$ in terms of the scaled sample matrix and applying the properties of unitary matrices, we have
\begin{align}
\begin{split}\notag
    \frac{1}{\sqrt{N-1}}\,D\,\frac{1}{\sqrt{N-1}}\,D^T= K & = U\,\Sigma\,V^T\,V\,\Sigma^T\,U^T, \quad V^T\,V = \mathbb{I} \\
    K & = U\Sigma\,\Sigma^T\,U^T \\
    K & = U\,\Sigma^2\,U^T
 \end{split}  \\
 K\,U & = U\,\Sigma^2 \label{eq:SVD_eigensystem}
\end{align}
The eigensystem of the covariance matrix $K$ given in Equation \ref{eq:SVD_eigensystem} is equivalent to the numerical approximation of the Fredholm integral equation when the quadrature weights are all equal to one (see \cite{safta2026kle}). Here, the eigenvalues ($\lambda_i , \; i=1,\cdots,p$) of $K$ are equivalent to the squares of the singular values ($\sigma_i$) of $D/\sqrt{n-1}$. Likewise, the right eigenvectors of $K$ correspond to the columns of $U$, which are composed of the left singular vectors of $D/\sqrt{n-1}$. The $k$th column of $U$ is denoted by $\mathbf{u}^{(k)}$. Using the eigenvectors of $K$, the eigenfunctions of the covariance function (of Equation~\ref{eq:integral}) can be obtained as
\begin{equation}
     \psi_{k}(\xi^1,\xi^2) = \boldsymbol{\phi}(\xi^1,\xi^2)^{t}\,\mathbf{u}^{(k)} \, ,\quad k =1,2,\cdots, p\,;\quad \mathbf{u}^{(k)} \in \mathbb{R}^{p} 
 \label{eq:eigenfunctions_SVD}
\end{equation}
The coefficients of the eigenfunctions in the KL expansion (Equation~\ref{eq:expansion}) are here referred to as the principal component scores (PC scores) or principal components. The PC scores can be computed by projecting the data on the eigenfunctions (obtained in Equation~\ref{eq:eigenfunctions_SVD}) as shown in Equation~\ref{eq:expansion}. 

The dimension of the system can be reduced by taking a finite number ($N_K < p$) of KL terms to approximate the expansion for each observation as:
\[ \mathcal{X}_i(\xi^1,\xi^2)  \approx \sum_{k=1}^{N_K}  \, \psi_k(\xi^1,\,\xi^2) \, \zeta_{k\,i}\qquad i =1,\cdots,N\]
This process results in a significant reduction in the dimensionality of the data by choosing an $N_K$ (number of eigenfunctions) much smaller than the size of the discretized samples of $\mathcal{X}(\xi^1,\xi^2)$. To recover the deformed configuration, we take  ${\Phi^1}_i\approx\mu^1(\xi^1,\xi^2)+\sum_{k=1}^{N_K}  \, \psi_k(\xi^1,\,\xi^2) \, \zeta_{k\,i}$. For the other two components of deformation, $\mathcal{Y}(\xi^1,\xi^2)$, and $\mathcal{Z}(\xi^1,\xi^2)$, FPCA is performed in the same manner as above. The PC scores thus obtained are used in a supervised learning framework as described in the next subsection. 

\begin{remark} 
A key advantage of FPCA is its ability to satisfy the geometric constraints described above exactly through the construction of an appropriate Karhunen-Loève (KL) basis. In contrast, neural network-based dimensionality-reduction methods, such as variational autoencoders (VAEs) \cite{kingma2022autoencodingvariationalbayes}, generally enforce these constraints only approximately. FPCA also offers a significant practical advantage over neural network-based autoencoders in determining the dimension of the latent representation. Autoencoders require the latent-space dimension to be specified a priori, and the reconstruction accuracy depends strongly on this choice. Consequently, the latent dimension becomes an important hyperparameter that must be tuned iteratively to achieve a desired level of variance retention. In contrast, FPCA determines the required number of modes, denoted by $K$, automatically from a prescribed variance-retention threshold. Thus, the user specifies the desired level of accuracy indirectly via the retained variance, and the latent-space dimension is inferred directly from the data based on the chosen threshold. These two properties---(i) exact satisfaction of the system’s geometric constraints through the selection of a specialized KL basis and (ii) automatic determination of the latent-space dimension based on a target variance threshold---make FPCA particularly well-suited for dimensionality reduction in the systems considered in this work, offering a desirable alternative over neural network-based autoencoders.
\end{remark}

\subsection{Supervised learning in the reduced dimension via deep neural network}\label{sec:NN}

We use a feedforward neural network model to learn the PC scores obtained via FPCA. This neural network was employed to map the three input parameters to the target functional principal component analysis (FPCA) coefficients. The network architecture consists of seven linear layers with a progressively expanding-then-contracting hidden layer configuration of widths $n$, $2n$, $2n$, $2n$, $4n$, and $n$ (with $n = 500$), interposed between the input layer of dimension 3 and the output layer of dimension equal to the number of retained FPCA components. For clarity, the Appendix includes a table summarizing these architecture details. Rectified linear unit (ReLU) activation functions were applied following each hidden layer, while the final output layer was left unactivated to permit unrestricted regression targets. The resulting network comprised 5,551,087 trainable parameters. Model weights were optimized using the Adam algorithm with an initial learning rate of $10^{-2}$ and an $L_2$ weight-decay coefficient of $10^{-4}$ to mitigate overfitting, in conjunction with a plateau-based learning rate scheduler (ReduceLROnPlateau) that reduced the learning rate by a factor of 0.75 upon stagnation of the validation loss over a patience window of 10 epochs, subject to a lower bound of $10^{-6}$. Training was performed for 1000 epochs with a mini-batch size of 32 using the mean squared error (MSE) loss function. Prior to training, the target FPCA coefficients were normalized via a modified z-score scaling scheme, wherein the standardized (zero-mean, unit-variance) coefficients were subsequently rescaled to the range $[0, 100]$, a transformation designed to stabilize gradient magnitudes during optimization and improve convergence behavior. The dataset was partitioned into training and validation subsets using a 90:10 split with a fixed random seed to ensure reproducibility, and model performance was assessed via the Pearson correlation coefficient between predicted and ground-truth values on the held-out validation datasets. Training, validation, and testing of the DNNs are conducted using PyTorch \cite{paszke2019pytorch}.

As previously discussed, the high-dimensional nature of $\Phi$ poses a challenge in approximating the map $\mathcal{H}$ (from the system parameters to the deformation $\Phi$) through a machine learning model. This is alleviated by the aforementioned dimensionality-reduction technique, which enables us to represent $\Phi$ through a finite number of PC scores, $\zeta_{k}, k=1,2,\cdots,N_K$. Therefore, we approximate the map $\mathcal{H}$ by another map $\bar{\mathcal{H}}$, $\bar{\mathcal{H}} : \{ N_w, m, \Theta\} \rightarrow \{\zeta_{k}\}_{k=1}^{N_K}$ in the latent space, and the ML model learns this new map.

The inputs to the neural network are also the system parameters, specifically the total number of graphene layers ($N_w$), the layer number $m$, and the applied strain denoted by $\Theta$. The labels/outputs of the neural network---the PC scores---are obtained from the 2D manifolds representing deformed graphene layers. A collection of 2D manifolds is obtained from the homogenized bulk multilayer graphene by discretizing the bulk multilayer graphene along the direction normal to the graphene layers (i.e., along $\xi^3$). These 2D manifolds are parametric surfaces of the ($\xi^1,  \xi^2$) plane, given as ${\Phi} 
\xrightarrow[\text{}]{\text{}}
   \left\{
     \Phi^1(\xi^1,\xi^2,\xi^3_m), 
     \Phi^2(\xi^1,\xi^2,\xi^3_m), 
     \Phi^3(\xi^1,\xi^2,\xi^3_m)
   \right\}$,
$m=1,\cdots,N_{w}$
where the $\xi^3$ coordinate of the $m^\text{th}$ layer in the parametric domain is given by $\xi^3_m$. Note that the deformation includes displacement due to applied strain. Thus, to facilitate the use of the Fourier basis, we subtract the non-periodic component before applying the Fourier transform. After training and inference in the latent space, we add the non-periodic component to the high-dimensional data to retrieve the total deformation.

This map $\bar{\mathcal{H}}$ can be approximated through any regression method, including neural networks, Gaussian process regression, etc. However, we choose to approximate this map using a deep neural network (DNN). DNNs possess high accuracy and versatility in predicting a wide variety of data. DNNs offer the additional advantage of simplicity compared to other methods, and the functionality is sufficient for our particular use case.

\section{Data Generation and Preprocessing}\label{sec:dat_gen_aug}

\subsection{Data generation}\label{sec:dat_gen}

The atomistic-continuum (foliation \cite{GhoshJMPS2013}) model is used to simulate multilayer graphene for data generation. Freely suspended multilayer graphene is subjected to three loading conditions: uniaxial compression, biaxial compression, and biaxial compression combined with simple shear. For each loading condition, the training dataset consists of systems with  $N_w=$ 2, 5, 7, 10, 12, 15, 17, and 20. We evaluate the model’s performance on three unknown systems, 4, 11, and 18 layer graphene. Each system is square with dimensions of $100\times N_w$ nm on each side. The simulations produce three-dimensional deformation fields, from which datasets of the corresponding two-dimensional deformed graphene surfaces are extracted. All simulations were performed on 64 cores of Superior, Michigan Technological University's high-performance computing facility, with nodes containing 32 CPU cores (Intel Xeon E5-2683 2.10 GHz) and 256 GB RAM. The total wall time to generate each training data set was 1131.28 hours for uniaxial compression, 2241.72 hours for biaxial compression, and 2003.81 hours for biaxial compression with simple shear, excluding the minimal amount of preprocessing time, which took less than 1\% of the total time to run for each loading case on a single core. These are one-time costs and do not increase regardless of the number of predictions made by the ML model.

When computing the principal components, FPCA is performed separately on the $x_1$-, $x_2$-, and $x_3$-direction deformations. The $x_1$- and $x_2$-direction deformations are each composed of two components: the displacement induced by the applied uniform compression and the displacement associated with the non-uniform wrinkling, which vanishes at the boundaries. To facilitate an efficient Fourier-based representation in FPCA, only the wrinkling-induced displacement field is used for dimensionality reduction. The displacement due to the applied uniform compression is subsequently added back to reconstruct the total in-plane displacement field.

\subsection{Data augmentation}\label{sec:dat_aug}

\begin{figure}[htbp]
    \centering
\begin{subfigure}[c]{0.99\textwidth}
        \centering
        \includegraphics[width=\textwidth]{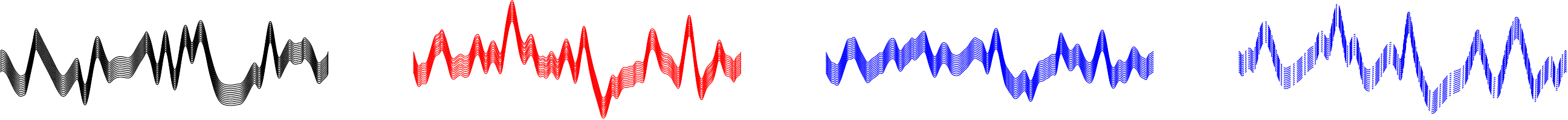}
    \end{subfigure}
    \hfill
    \begin{subfigure}[c]{0.22\textwidth}
        \centering
        \caption{}
    \end{subfigure}
    \hfill
    \begin{subfigure}[c]{0.24\textwidth}
        \centering
        \caption{}
    \end{subfigure}
    \hfill
    \begin{subfigure}[c]{0.24 \textwidth}
        \centering
        \caption{}
    \end{subfigure}
    \hfill
    \begin{subfigure}[c]{0.22\textwidth}
        \centering
        \caption{}
    \end{subfigure}
    \caption{Cross-sections of out-of-plane deformations of 12-layer graphene results at 1\% biaxial compression from a) the original foliation code, b) the ML prediction without data augmentation, c) the ML prediction with data augmentation, and d) the minimized ML prediction with augmentation. While the shape differs, none of the ML predictions appear inherently worse than another or significantly different from the original.}
    \label{fig:ils_layers}
\end{figure}

The size of the dataset is inherently constrained by the computational cost of running the atomistic-continuum simulations for multiple configurations. Consequently, the dataset used in this work is relatively modest, with the training set composed of only eight multilayer graphene configurations containing between 2 and 20 layers. While this dataset is sufficient to accurately learn the in-plane deformation of the graphene sheets, it is inadequate for predicting the inter-layer spacing with comparable accuracy, which can impair convergence during energy minimization. This limitation arises because the repulsive component of the non-bonded interaction, modeled as the inverse tenth power of the inter-layer spacing, becomes extremely sensitive when adjacent layers approach one another. As a result, underpredictions of the inter-layer spacing generate disproportionately large repulsive forces, which impair the convergence of the foliation model. Consequently, the energy minimization process is highly sensitive to prediction errors in the inter-layer spacing. To mitigate this, we employ data augmentation on the system. When assessing the effectiveness of augmentation, its necessity is not immediately evident from the appearance of the cross-sections. As seen in the cross-sections of 12-layer graphene under 1\% biaxial compression in Figure \ref{fig:ils_layers}, the predictions, both augmented and non-augmented, do not display a particularly different morphology compared to the original on the left. (Note that 12-layer graphene was initially an unknown system during the exploration of data augmentation but was subsequently incorporated into the training dataset.)

\begin{figure}[htbp]
    \centering
    \begin{subfigure}[c]{0.99\textwidth}
        \centering
        \includegraphics[width=\textwidth]{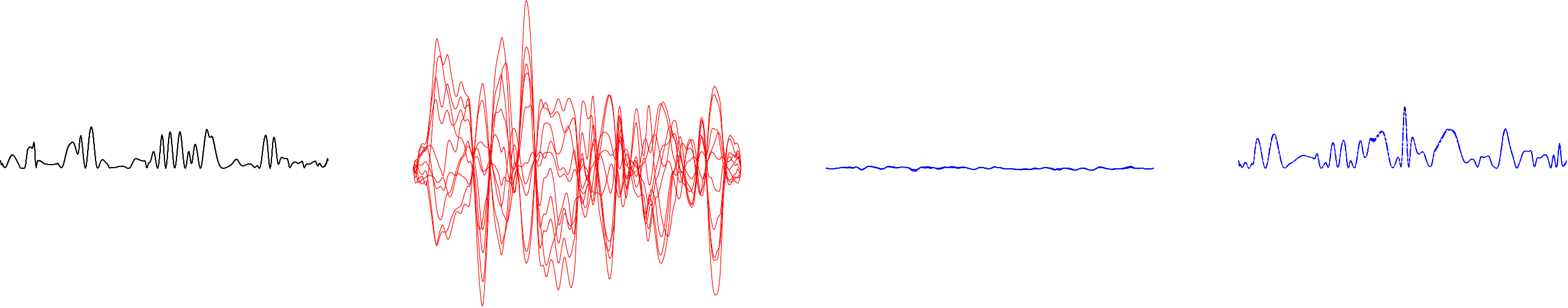}
    \end{subfigure}
    \hfill
    \begin{subfigure}[c]{0.22\textwidth}
        \centering
        \caption{}
    \end{subfigure}
    \hfill
    \begin{subfigure}[c]{0.24\textwidth}
        \centering
        \caption{}
    \end{subfigure}
    \hfill
    \begin{subfigure}[c]{0.24 \textwidth}
        \centering
        \caption{}
    \end{subfigure}
    \hfill
    \begin{subfigure}[c]{0.22\textwidth}
        \centering
        \caption{}
    \end{subfigure}
    \caption{The layer spacing (transversal stretch) from the above cross-sections of deformation. a) The layer spacing from the original foliation code is largely uniform between layers and shows that certain regions have greater layer spacing, loosely correlated with wrinkle locations. b) Without data augmentation, layer spacing of the prediction is erratic and of abnormally high magnitude in some places, and it is not at all uniform between layers throughout the sample, leading the energy minimization to fail. c) Layer spacing is highly uniform with much less variation than the original. This prediction is better suited to the energy minimization. d) After minimizing the ML prediction, the layer spacing profile appears realistically similar to the original foliation code results.}
    \label{fig:ils_spacing}
\end{figure}

However, when attempting to provide the predictions to the foliation code for minimization, the non-augmented data fails to minimize. By inspecting the interlayer spacing, represented via transversal stretch \(\lambda^t\), we find the issue becomes especially apparent in the non-augmented ML prediction versus the original (Figure~\ref{fig:ils_spacing}(b) and (a), respectively). The prediction exhibits aberrantly variable layer spacing, leading to failure of energy minimization due to the aforementioned reason. To overcome this challenge, we augment the data by interpolating the deformation manifolds between actual layers of the system to obtain interpolated layers. While these interpolated layers do not exist physically, they enable the ML model to better capture the variation between layers and avoid producing results with erroneous layer spacing. Examining the resulting layer spacing of the prediction using augmented data (Figure~\ref{fig:ils_spacing}(c)) clearly showcases the advantage of using the augmented dataset for training. The elimination of nuance by the ML model is noteworthy, with the new prediction having much lower variation in layer spacing than the original; however, since the prediction is used as an initial guess and minimized in the foliation code, minimal variation is much preferable over erratic layer spacing in terms of computing energy. The foliation model is used to minimize the ML prediction, and the resulting ML-foliation prediction results are shown in Figures Figure~\ref{fig:ils_layers}(d) and Figure~\ref{fig:ils_spacing}(d), with the layer spacing profile being much more comparable to that of the foliation prediction.

\section{Results} \label{sec:results}

This section discusses the testing of the proposed deformation manifold learning model applied to suspended multilayer graphene under three different boundary conditions: uniaxial compression, biaxial compression, and combined biaxial compression with shear. To exactly satisfy the homogeneous displacement boundary conditions of the multilayer graphene simulations by the KL-expansion, these conditions are encoded into the Fourier basis sets. The presently adopted Fourier basis set encodes the geometric constraint of the deformation of multilayer graphene as $\phi_l(\xi^1 = \{0, 1\},\xi^2=\{0, 1\})=0;\; l = 1, 2, \cdots, p$; whenever the displacement is zero at the boundary. As a result, every function in the function space spanned by the Fourier basis set exactly satisfies the boundary conditions.

We perform FPCA separately on each displacement component and retain only the dominant principal components that capture $99.9\%$ of the cumulative variance in the training data. Owing to the distinct deformation morphologies associated with different loading conditions, a separate neural network is trained for each load case. However, for a given load case, a single network predicts the PC scores for all displacement components.

We reiterate that the wrinkling patterns are sensitive to the random initial perturbations applied to initiate wrinkling during simulations using the foliation model. Consequently, even for the same system subjected to identical loading conditions, the exact displacement field may vary from one simulation to another while exhibiting similar energy. This indicates the existence of multiple energetically similar equilibrium configurations, a well-established phenomenon for these systems \cite{ZHANGArroyo2014JMPS,Arroyo2004}). Nevertheless, the overall morphology of the wrinkling patterns remains consistent across different realizations. Accordingly, our objective is not to reproduce the exact displacement field of a particular simulation, but rather to accurately capture the characteristic wrinkle morphology while predicting a configuration that converges to the correct minimized energy state. This is accomplished by using the ML-predicted deformation as the initial guess for the foliation model, which subsequently performs energy minimization to obtain the corresponding atomistic-continuum equilibrium solution. The resulting solutions are hereafter referred to as those obtained using the ML-foliation method. The ML-foliation method is validated on unseen $4$, $11$, and $18$-layer graphene systems for three load cases. The energy minimization for both the ML-foliation and the foliation model was performed on the same HPC infrastructure using the same computational resources, with 64 CPU cores allocated to each calculation.

\subsection{Uniaxial Compression}
\begin{figure}[htbp]
    \centering
    \includegraphics[width=0.44\linewidth]{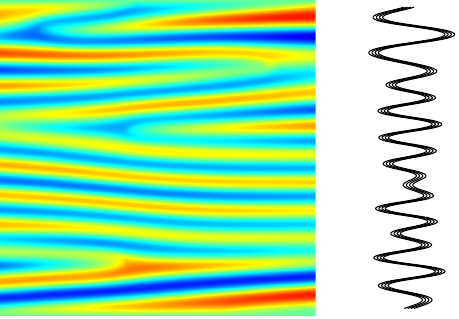}
    \includegraphics[width=0.549\linewidth]{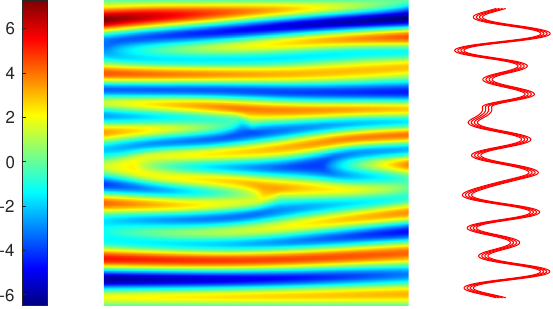}
    \caption{4-layer graphene under 5\% uniaxial compression: out-of-plane deformation of the top layer predicted by the foliation model (left) and ML-foliation model (right), including central cross-sections in the $x_2$-$x_3$ plane.}
    \label{fig:uniax_4_def}
\end{figure}

Multilayer graphene subjected to uniaxial compression exhibits predominantly sinusoidal wrinkling patterns with occasional bifurcations, as shown in Figure \ref{fig:uniax_4_def}. Due to the unconstrained boundaries parallel to the loading direction, the resulting deformation fields are non-periodic in the transverse direction. Consequently, direct representation of these deformations using a finite Fourier basis may introduce approximation errors. To address this limitation, a modified deformation representation is introduced in which the displacement field smoothly decays to zero outside the physical domain. This treatment improves the accuracy of the Fourier-based representation and enhances the performance of the machine-learning surrogate, leading to reduced prediction errors after subsequent energy minimization within the foliation framework.

\begin{figure}[htbp]
    \centering
    \includegraphics[width=0.32\linewidth]{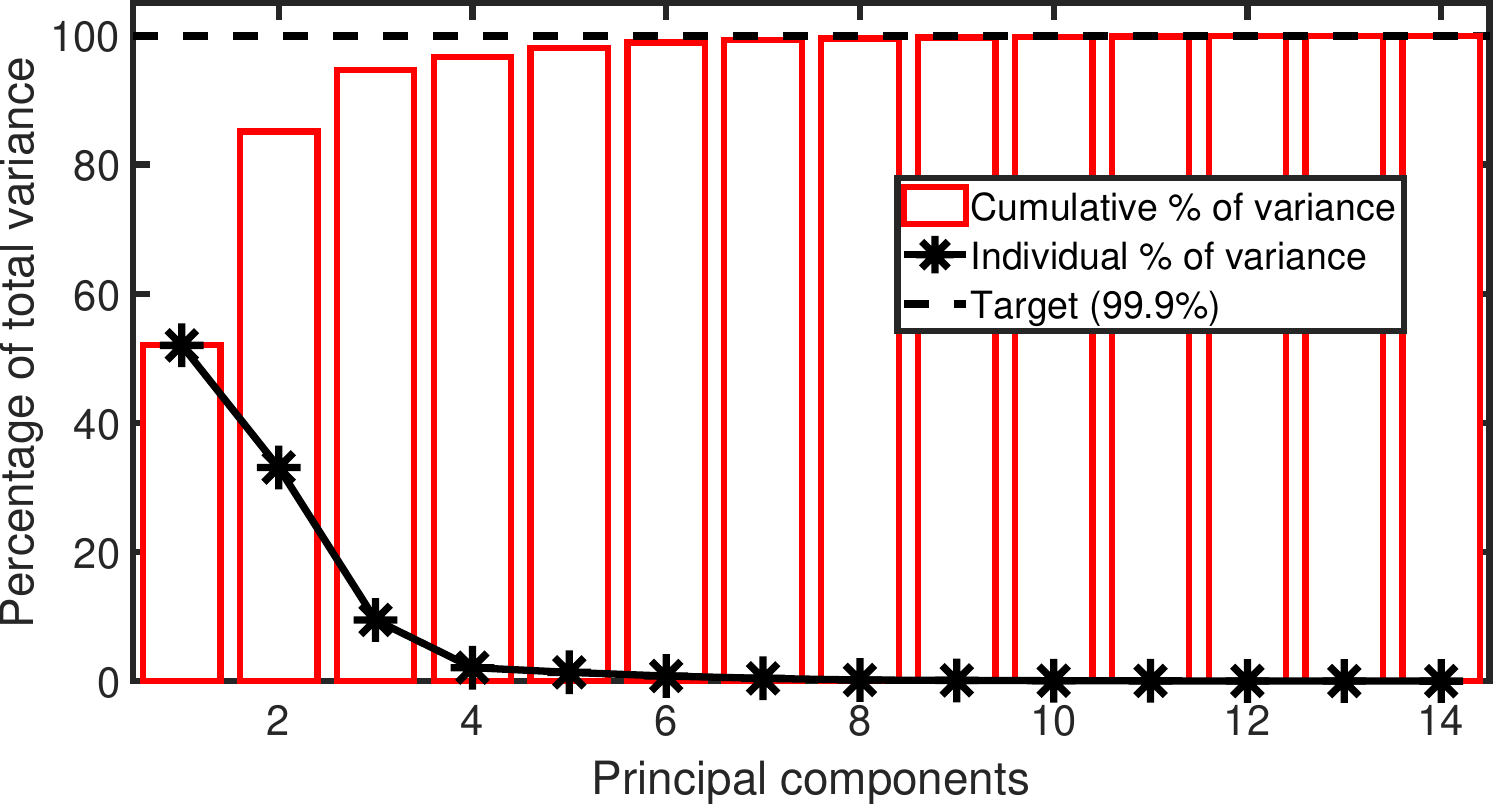}
    \includegraphics[width=0.32\linewidth]{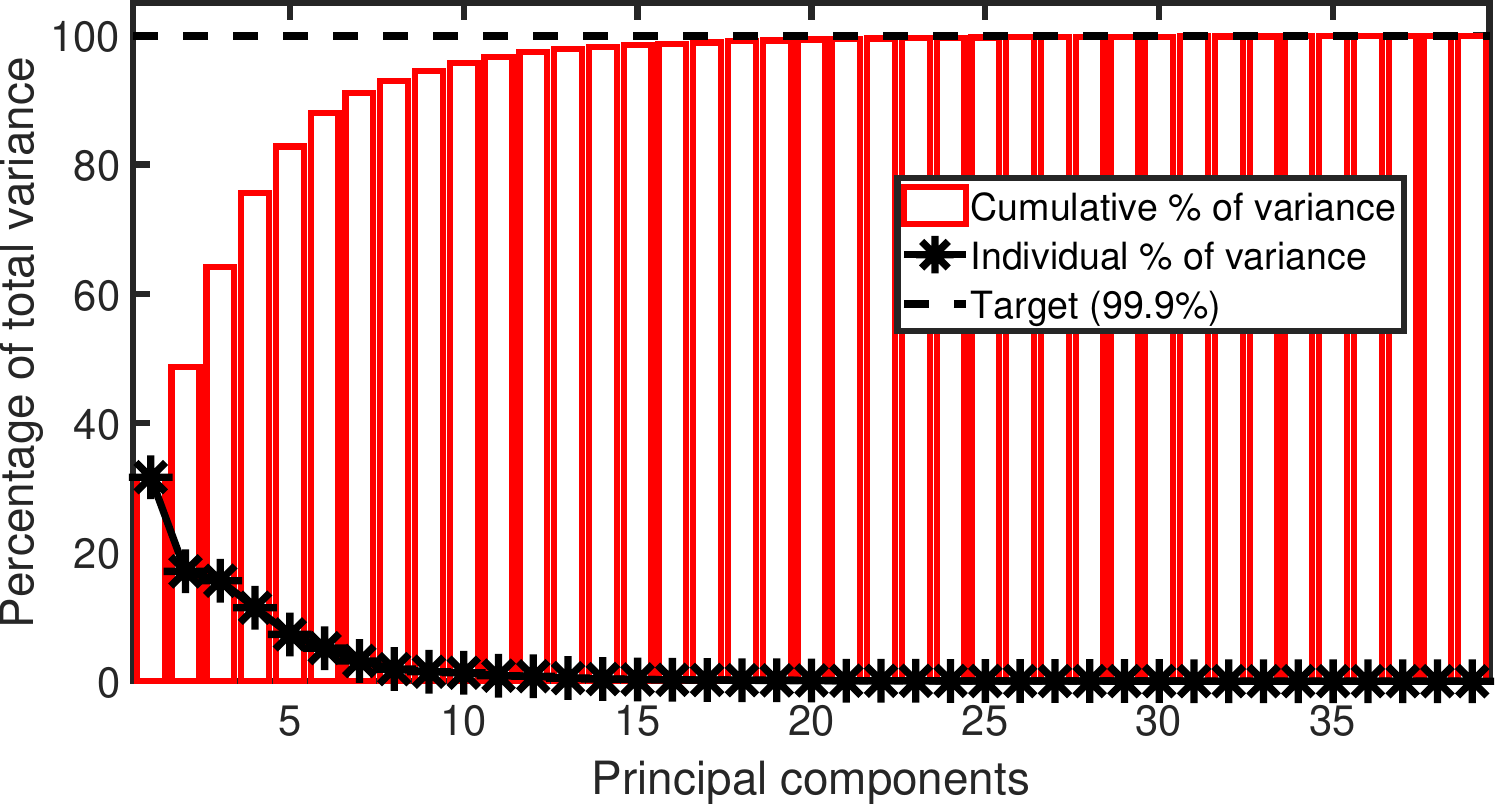}
    \includegraphics[width=0.32\linewidth]{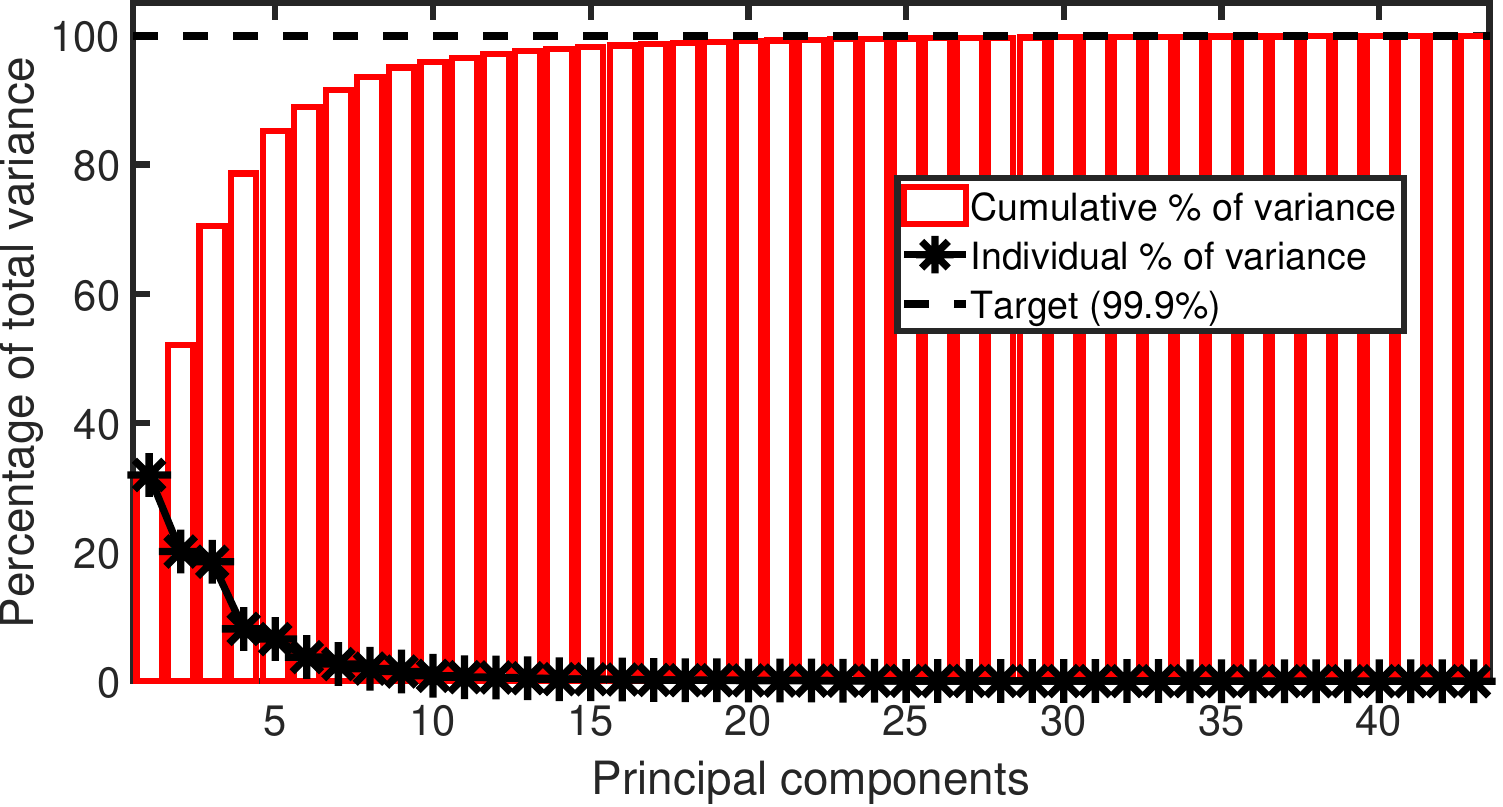}
    \caption{Scree and cumulative variance plots for FPCA of the uniaxial compression loading case. The horizontal axis represents the order of the principal components, while the vertical axis shows the cumulative and individual percentages of variance captured by each principal component, represented by the red bars and black lines, respectively. The three plots from left to right correspond to the $x_1$, $x_2$, and $x_3$  deformation components. The total number of principal components included in each plot is the number required to represent $99.9\%$ cumulative variance, as indicated by the black dashed line.}
\label{fig:kle_variance_uniax_gpn}
\end{figure}

Functional principal component analysis (FPCA) is applied to the deformation data obtained from uniaxially compressed multilayer graphene to quantify the contribution of the primary deformation modes. As shown in the scree plots in Figure~\ref{fig:kle_variance_uniax_gpn}, 14, 39, and 43 principal components are sufficient to capture $99.9\%$ of the variance in the $x_1$, $x_2$, and $x_3$ deformation components, respectively. Given that the largest system (consisting of $20$ layers) is simulated using $4.8$ million degrees of freedom (DOFs), and that each layer is originally represented by up to approximately $0.75$ million DOFs, FPCA reduces the dimensionality by approximately four orders of magnitude while retaining  $99.9\%$ of the variance in the data. Furthermore, the first principal component alone captures $31-52\%$ of the total variance for the different deformation components, indicating that the deformation response is governed primarily by a small number of dominant eigenfunctions. The neural network model trained to predict the corresponding PC scores achieves a Pearson correlation coefficient greater than $0.999$ on the validation dataset.

\begin{figure}[htbp]
    \centering
    \includegraphics[width=0.3\linewidth]{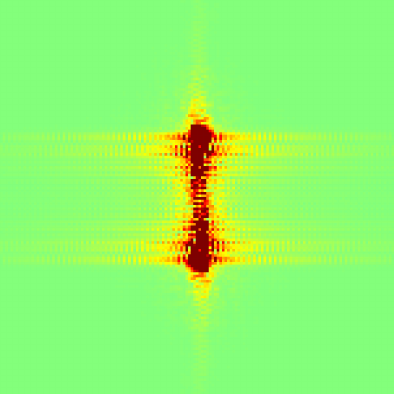}
    \includegraphics[width=0.425\linewidth]{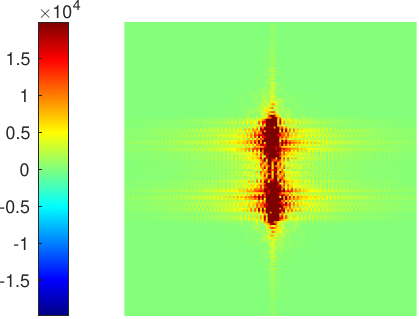}
    \caption{4-layer graphene under 5\% uniaxial compression: Comparison of 2D Fourier amplitude spectra of the out-of-plane deformation obtained by the foliation model (left) and the ML-foliation model (right).}
\label{fig:uniax_4_fft_comp_img}
\end{figure}

The first test case considers a 4-layer graphene system subjected to uniaxial compression. The deformation fields of the top layer predicted by the foliation model and the ML-foliation model are compared in Figure~\ref{fig:uniax_4_def}. The corresponding cross-sections through the middle of the sheet are also presented. Although the predicted deformations are not identical, they exhibit similar wrinkling patterns. The cross-section views also offer a comparison of the wavelength and other features, showing that there is a noticeable discrepancy in wavelength, but also that the ML-foliation prediction captures one of the unique features from the foliation prediction, in this case, a double-peak feature. This agreement is further confirmed by comparing the Fourier spectra of the out-of-plane deformation fields (Figure~\ref{fig:uniax_4_fft_comp_img}), where the two methods exhibit similar spectral distributions. These results demonstrate that the ML-foliation model accurately captures the overarching spatial characteristics of the wrinkling patterns. 

\begin{figure}[htbp]
    \centering
    \includegraphics[width=0.35\linewidth]{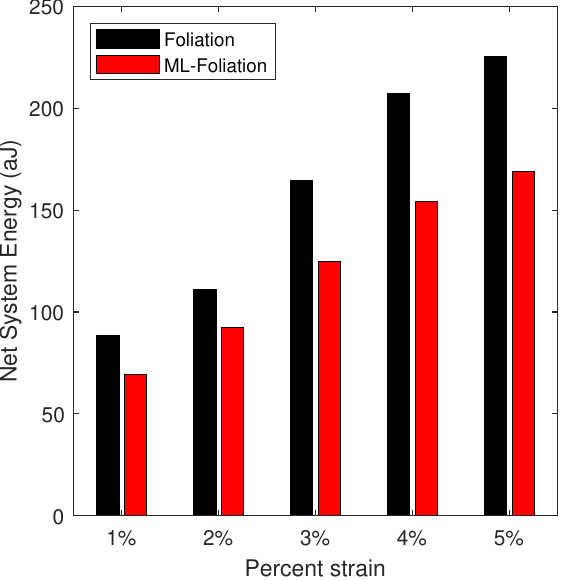}
    \includegraphics[width=0.35\linewidth]{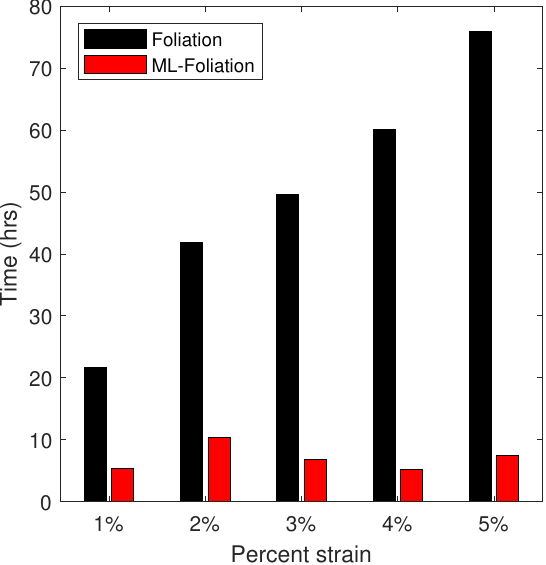}
    \caption{4-layer graphene under uniaxial compression: Energy (left) and time (right) comparison for the foliation model versus the ML-foliation model, denoted by black and red bars, respectively.}
    \label{fig:uniax_4_ener_time_bars}
\end{figure}

The energies and computational costs associated with the two approaches are compared in Figure \ref{fig:uniax_4_ener_time_bars}. The ML-foliation model converges to configurations with comparable or lower energies than those obtained using the foliation method. The underlying reason for this phenomenon is not apparent; however, it may be attributed to the improved initialization provided by the ML model, which learns deformation patterns from multiple similar simulations. The reduced wavelength in the ML-foliation prediction would also partially contribute to lower energy values. 
In addition, the ML-foliation approach significantly reduces computational time, achieving speedups ranging from 4.0$\times$ to 11.4$\times$ depending on the applied strain level.

\begin{figure}[htbp]
    \centering
    \includegraphics[width=0.44\linewidth]{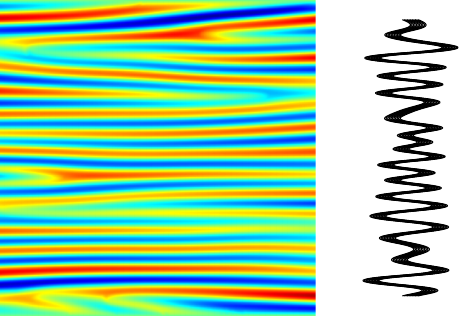}
    \includegraphics[width=0.545\linewidth]{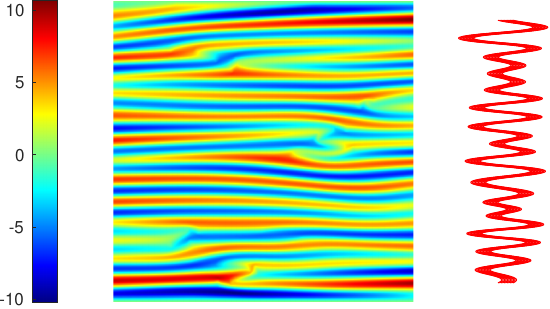}
    \caption{11-layer graphene under 5\% uniaxial compression: out-of-plane deformation of the top layer predicted by the foliation model (left) and ML-foliation model (right), including central cross-sections in the $x_2$-$x_3$ plane.}
    \label{fig:uniax_11_def}
\end{figure}

\begin{figure}[htbp]
    \centering
    \includegraphics[width=0.3\linewidth]{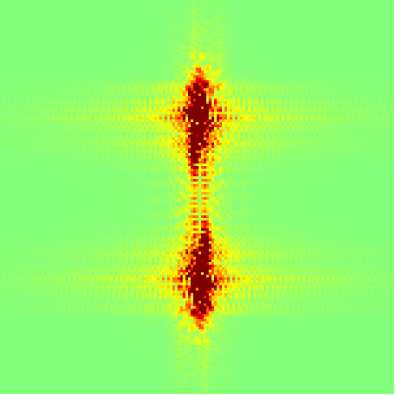}
    \includegraphics[width=0.41\linewidth]{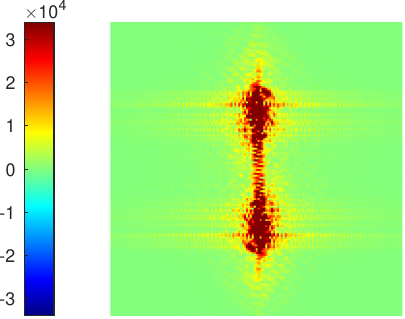}
    \caption{11-layer graphene under 5\% uniaxial compression: Comparison of 2D Fourier amplitude spectra of the out-of-plane deformation obtained by the foliation model (left) and the ML-foliation model (right).}
\label{fig:uniax_11_fft_comp_img}
\end{figure}

\begin{figure}[htbp]
    \centering
    \includegraphics[width=0.35\linewidth]{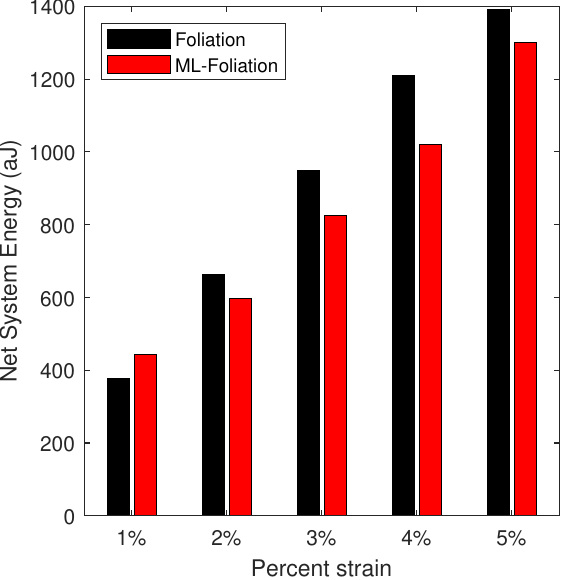}
    \includegraphics[width=0.35\linewidth]{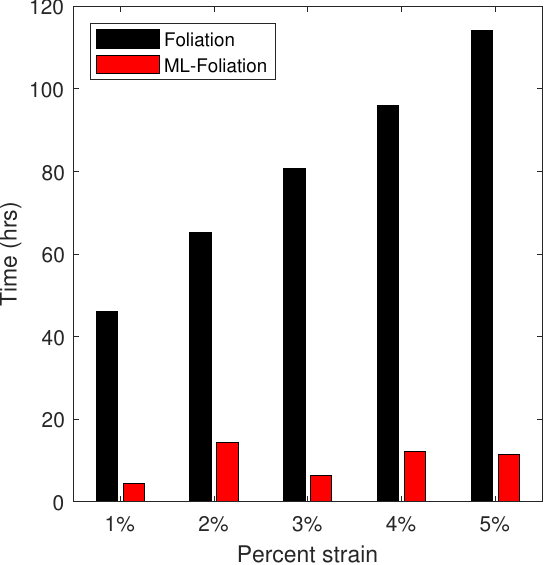}
    \caption{11-layer graphene under 5\% uniaxial compression: Energy and time comparison for foliation model and ML-foliation model, denoted by black and red bars, respectively. }
    \label{fig:uniax_11_ener_time_bars}
\end{figure}

The second test case considers an 11-layer graphene system, representing another unseen configuration for the model. The out-of-plane deformation fields under $5\%$ uniaxial compression are shown in Figure~\ref{fig:uniax_11_def}. The comparison indicates that the wrinkling morphology reproduced by the ML-foliation model effectively captures the deformation characteristics from the foliation model. The cross-sections also indicate high similarity in the wavelength. The 11-layer graphene system exhibits a greater number of wrinkles than the 4-layer graphene system. At higher bending angles (and, consequently, higher strain), the bending stiffness of multilayer graphene increases with the number of layers at a much lower rate than expected due to slip-induced softening (see \cite{han2020FLGslip}), whereas the system size increases $11/4$-fold in each direction, resulting in a greater number of wrinkles in the specimen. The Fourier spectra of the out-of-plane deformation at $5\%$ compression (Figure~\ref{fig:uniax_11_fft_comp_img}) exhibit similar distributions for both methods, further demonstrating that the ML-foliation approach accurately captures the dominant features of the deformation. The predicted energies and computational times are compared in Figure~\ref{fig:uniax_11_ener_time_bars}. The energies obtained from the ML-foliation model are generally consistent with those from the foliation model and are typically slightly lower, following a similar trend observed for the 4-layer graphene system. Meanwhile, the ML-foliation approach achieves substantially faster convergence, with computational speedups ranging from $4.6\times$ to $12.6\times$ compared to the foliation approach. The lowest strains will intuitively have reduced computational advantage since the foliation model costs less to simulate the system from the initial, undeformed configuration loaded to the target strain increment.

\begin{figure}[htbp]
    \centering
    \includegraphics[width=0.44\linewidth]{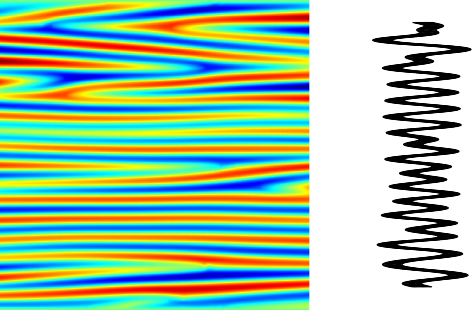}
    \includegraphics[width=0.545\linewidth]{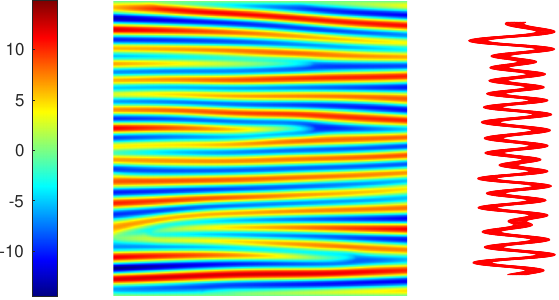}
    \caption{18-layer graphene under 5\% uniaxial compression: out-of-plane deformation of the top layer predicted by the foliation model (left) and ML-foliation model (right), including central cross-sections in the $x_2$-$x_3$ plane.}
    \label{fig:uniax_18_def}
\end{figure}

\begin{figure}[htbp]
    \centering
    \includegraphics[width=0.3\linewidth]{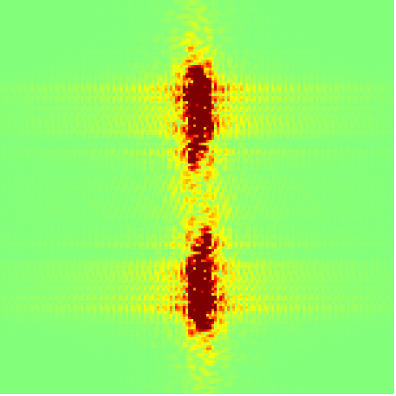}
    \includegraphics[width=0.41\linewidth]{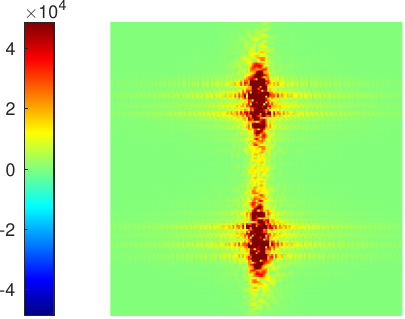}
    \caption{18-layer graphene under 5\% uniaxial compression: Comparison of 2D Fourier amplitude spectra of the out-of-plane deformation obtained by the foliation model (left) and the ML-foliation model (right).}
\label{fig:uniax_18_fft_comp_img}
\end{figure}

\begin{figure}[htbp]
    \centering
    \includegraphics[width=0.35\linewidth]{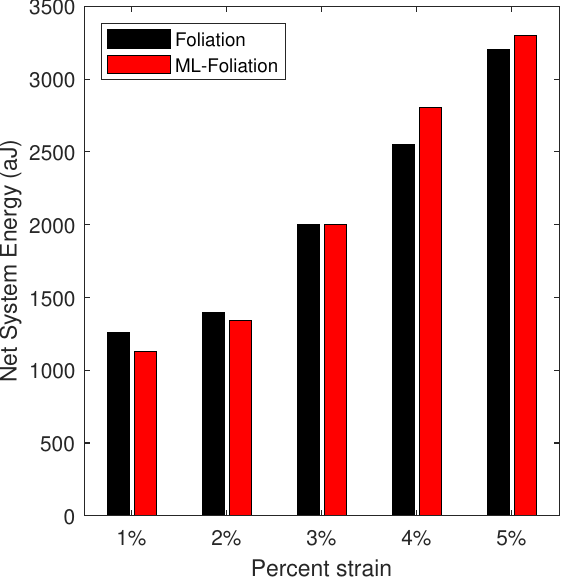}
    \includegraphics[width=0.35\linewidth]{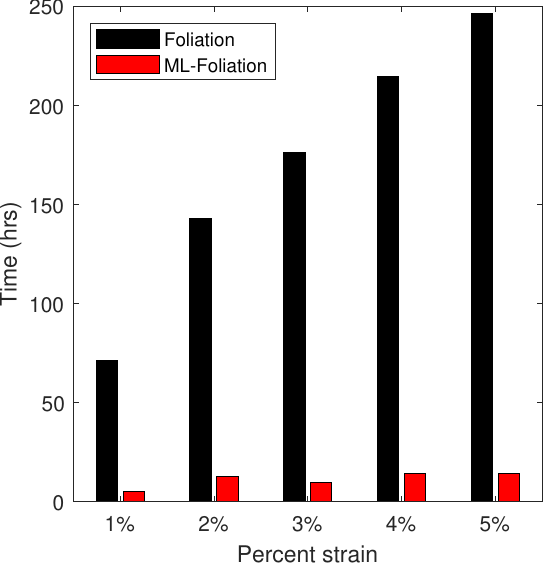}
    \caption{18-layer graphene under 5\% uniaxial compression: Energy and time comparison for foliation model and ML-foliation model, denoted by black and red bars, respectively.}
    \label{fig:uniax_18_ener_time_bars}
\end{figure}

The last uniaxial compression test case considers 18-layer graphene. The out-of-plane deformation fields at $5\%$ strain are presented in Figure~\ref{fig:uniax_18_def}. The cross-sections are also shown. Comparing the two approaches once again shows that the ML-foliation prediction captures many characteristics of the morphology from the foliation prediction. The cross-sections also indicate strong agreement in terms of the wavelength. The morphology match is confirmed by comparison of the Fourier spectra in Figure~\ref{fig:uniax_18_fft_comp_img}. The predicted energies and computation times are shown in Figure~\ref{fig:uniax_18_ener_time_bars}. Once again, the energies predicted by the ML-foliation model tend to be close to or lower than those predicted by the foliation model. Meanwhile, the ML-foliation continues to outperform the foliation model in terms of computational efficiency, with speedups ranging from $11.1\times$ to $18.1\times$, varying with strain.

\subsection{Biaxial Compression}

Biaxial compression is a more constrained loading condition, with strain applied to all four sides of the multilayer graphene sheet rather than just two. Due to the constrained nature of the loading, the deformations exhibit greater complexity than for uniaxial compression, with more junctions and terminations in the middle of each layer. This greater complexity elevates the computational cost of the simulations. Consequently, the largest systems become computationally prohibitive to run up to 5\% compression given our hardware constraints. Thus, the biaxial compression training data consists of manifolds only up to 4\% biaxial compression.

\begin{figure}[htbp]
    \centering
    \includegraphics[width=0.32\linewidth]{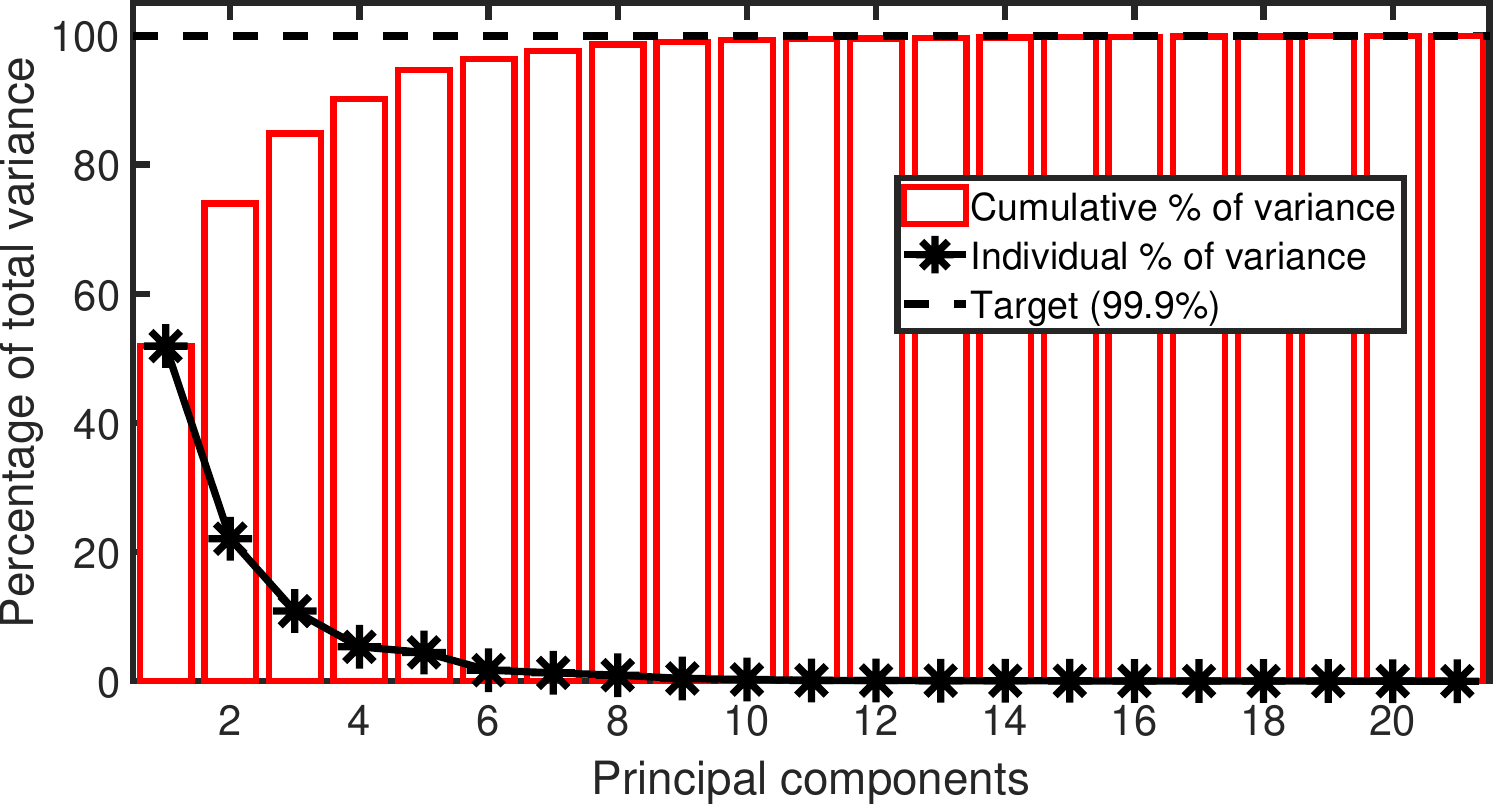}
    \includegraphics[width=0.32\linewidth]{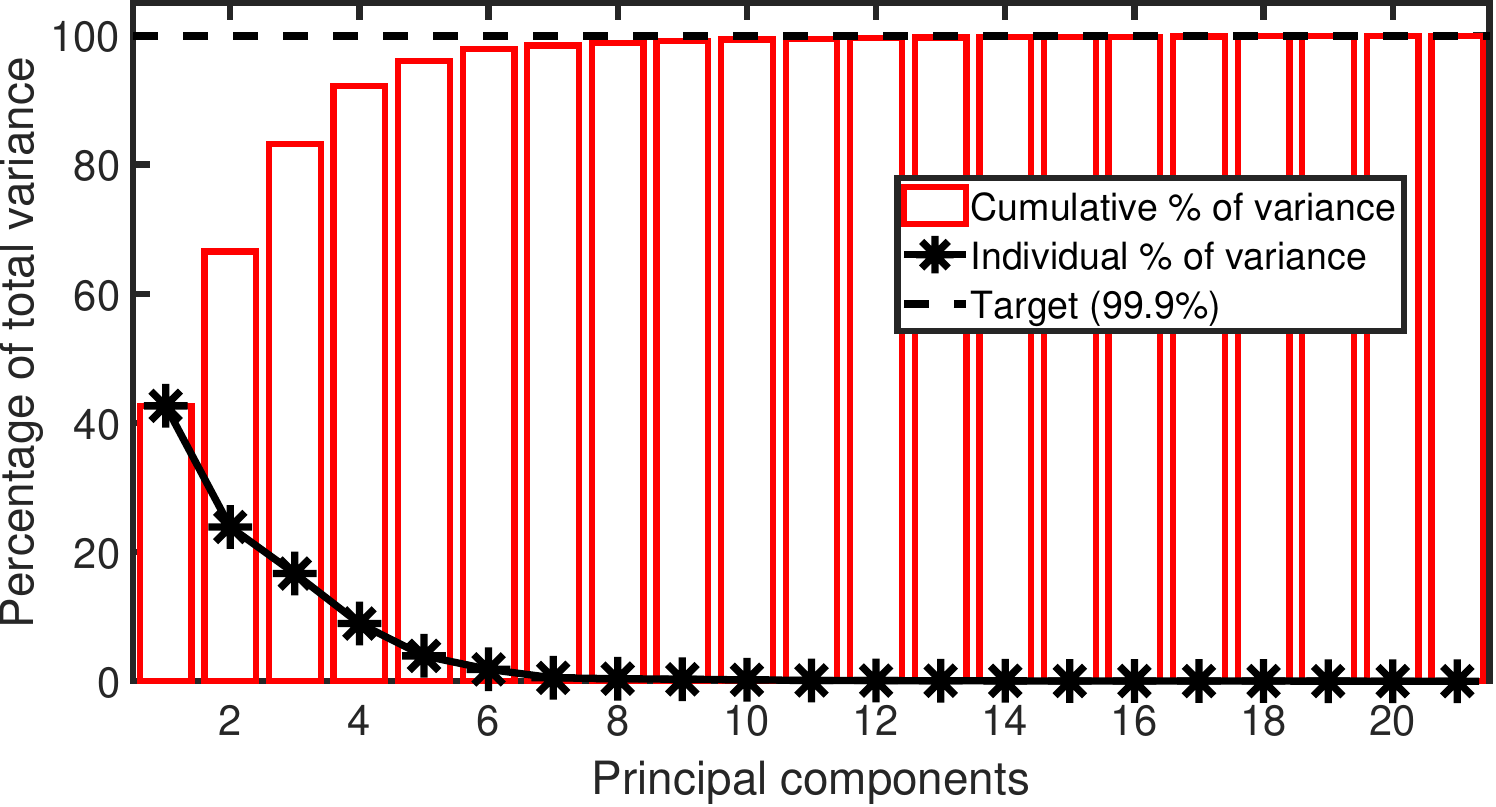}
    \includegraphics[width=0.32\linewidth]{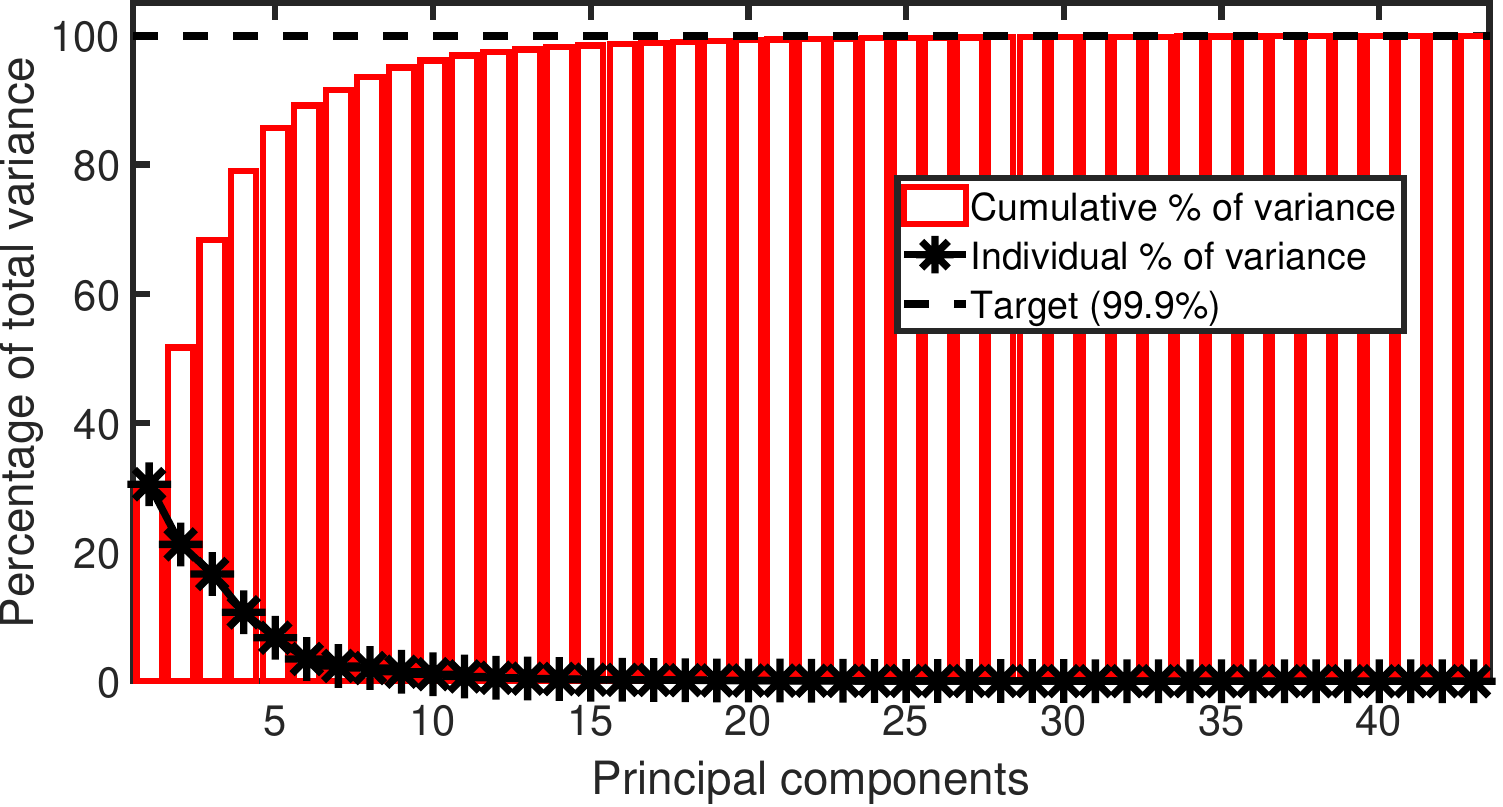}
    \caption{Scree and cumulative variance plots for FPCA for the biaxial compression loading case. The horizontal axis represents the order of the principal components, while the vertical axis shows the cumulative and individual percentages of variance captured by each principal component, represented via red bars and black lines, respectively. The three plots from left to right correspond to the $x_1$, $x_2$, and $x_3$ deformations. The total number of principal components included in each plot is the number required to represent 99.9\% cumulative variance, as indicated by the black dashed line.}
    \label{fig:kle_variance_biax_gpn}
\end{figure}

As with uniaxial compression, FPCA is performed on the deformation data obtained from simulating biaxial compression of multilayer graphene systems. The contribution of each eigenfunction is presented in the scree plots in Figure \ref{fig:kle_variance_biax_gpn}. For this load case, 21, 21, and 43 principal components suffice to capture 99.9\% of the variance in the $x_1$, $x_2$, and $x_3$ deformation components, respectively. Here, FPCA continues to demonstrate a significant reduction in dimensionality of approximately 4 orders of magnitude. Again, the first principal component in each direction captures a large share of the variance, from $30\%-52\%$ of the total variance, with the following principal components contributing progressively less, indicating that an accurate representation of the data is achieved through a relatively low number of eigenfunctions. One can also observe that the $x_1$ and $x_2$-deformations each require a similar number of principal components to reach $99.9\%$ variance. This is expected given that the strain is applied equally in each direction, leading to wrinkling patterns with similar complexity in either direction. Once again, the neural network trained on the PC scores demonstrates a Pearson correlation coefficient over 0.999 on the validation data.

\begin{figure}[htbp]
    \centering
    \includegraphics[width=0.44\linewidth]{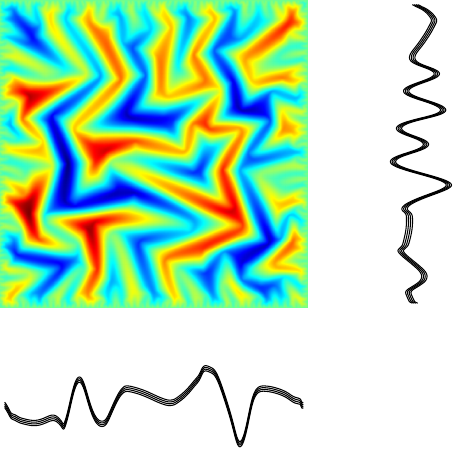}
    \includegraphics[width=0.553\linewidth]{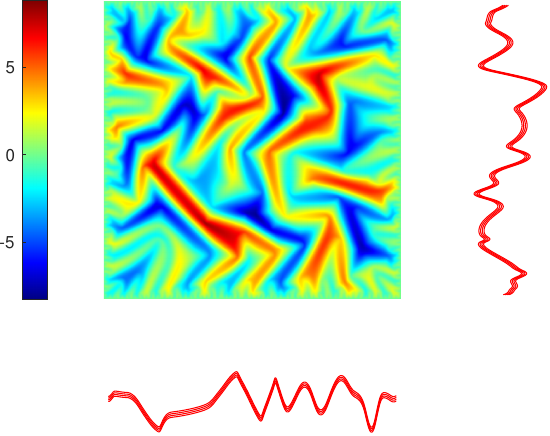}
    \caption{4-layer graphene under 3\% biaxial compression: out-of-plane deformation of the top layer predicted by the foliation model (left) and ML-foliation model (right), including central cross-sections, with the cross-sections to the right being in the $x_2$-$x_3$ plane, and the lower cross-sections being in the $x_1$-$x_3$ plane.}
    \label{fig:biax_4_3pc_orig_def}
\end{figure}

\begin{figure}[htbp]
    \centering
    \includegraphics[width=0.3\linewidth]{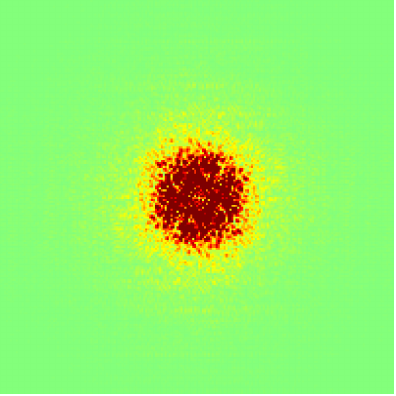}
    \includegraphics[width=0.41\linewidth]{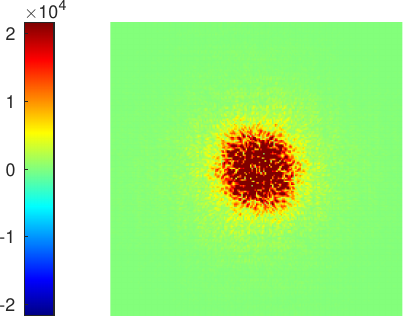}
    \caption{4-layer graphene under 3\% biaxial compression: Comparison of 2D Fourier amplitude spectra of the out-of-plane deformation obtained by the foliation model (left) and the ML-foliation model (right).}
\label{fig:biax_4_3pc_fft_comp_img}
\end{figure}

The first test of this load case considers 4-layer graphene under biaxial compression. Figure \ref{fig:biax_4_3pc_orig_def} presents the foliation and ML-foliation predictions for the out-of-plane deformations of the top layer of 4-layer graphene under $3\%$ strain, along with the cross-sections at the center in the $x_1$-$x_3$ and $x_2$-$x_3$ planes. The similarity in features and cross-sectional profiles suggests that the two sets of results possess similar morphology, a conclusion further supported by the comparison of the 2D Fourier spectra (Figure \ref{fig:biax_4_3pc_fft_comp_img}). 

\begin{figure}[htbp]
    \centering
    \includegraphics[width=0.35\linewidth]{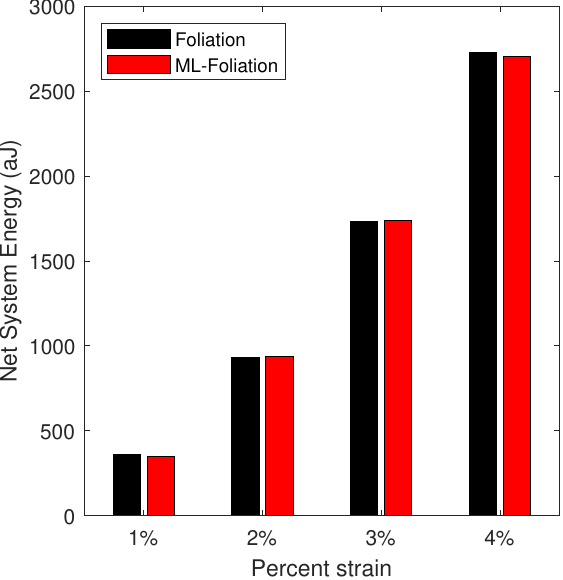}
    \includegraphics[width=0.35\linewidth]{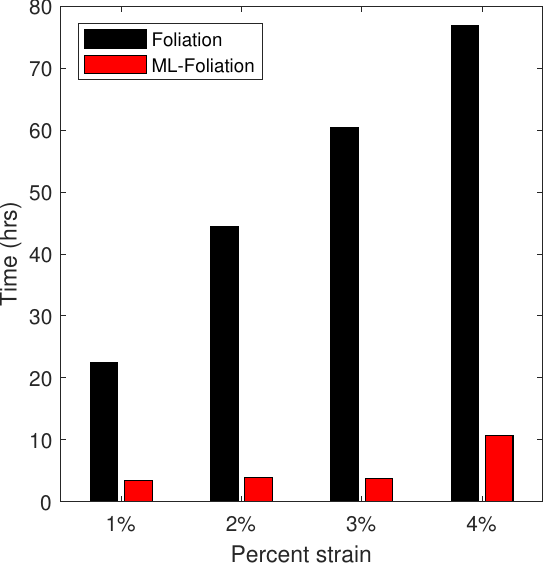}
    \caption{4-layer graphene under biaxial compression: Energy (left) and time (right) comparison for the foliation model and ML-foliation model, denoted by black and red bars, respectively.}
    \label{fig:biax_4_ener_time_bars}
\end{figure}

The energy and time metrics for 4-layer graphene under biaxial compression are presented here. Figure \ref{fig:biax_4_ener_time_bars} indicates a strong energy match, with the ML-foliation values falling nearer to the foliation results than for 4-layer graphene under uniaxial compression. While the ML-foliation predictions for uniaxial compression may have converged to more energetically desirable configurations, the deformation complexity of biaxial compression may have inhibited such a phenomenon, so similarity in energy values remains a strong indicator of realistic results. In addition, the ML-foliation model achieves speedups from $6.5\times$ to $16.0\times$, varying across strain increments. 

\begin{figure}[htbp]
    \centering
    \includegraphics[width=0.442\linewidth]{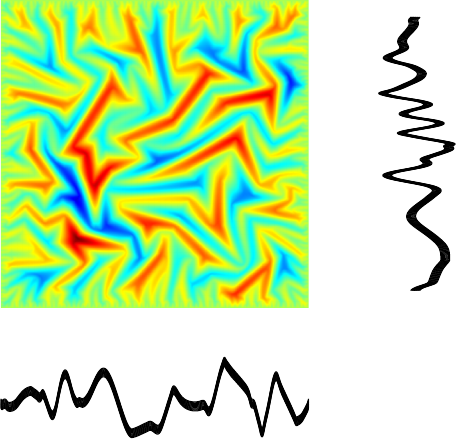}
    \includegraphics[width=0.54\linewidth]{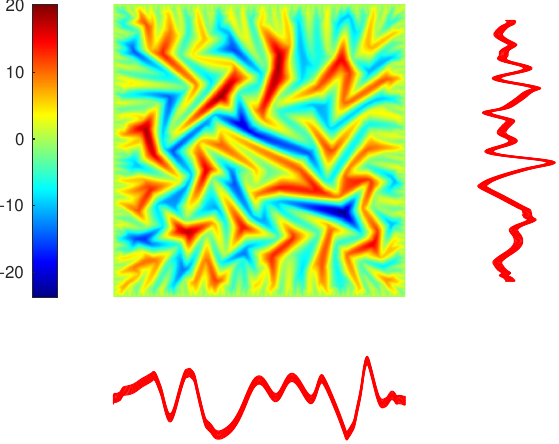}
    \caption{11-layer graphene under 3\% biaxial compression: out-of-plane deformation of the top layer predicted by the foliation model (left) and ML-foliation model (right), including central cross-sections, with the cross-sections to the right being in the $x_2$-$x_3$ plane, and the lower cross-sections being in the $x_1$-$x_3$ plane.}
    \label{fig:biax_11_3pc_orig_def}
\end{figure}

\begin{figure}[htbp]
    \centering
    \includegraphics[width=0.3\linewidth]{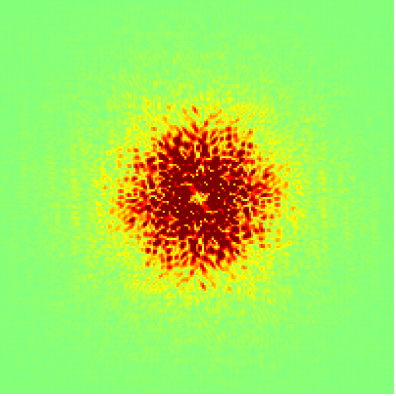}
    \includegraphics[width=0.41\linewidth]{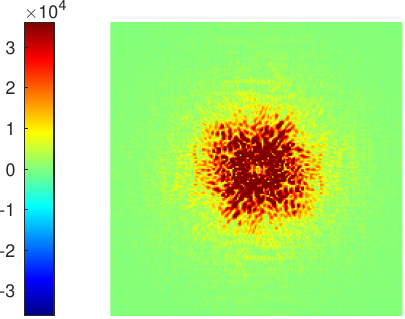}
    \caption{11-layer graphene under 3\% biaxial compression: Comparison of 2D Fourier amplitude spectra of the out-of-plane deformation obtained by the foliation model (left) and the ML-foliation model (right).}
\label{fig:biax_11_3pc_fft_comp_img}
\end{figure}

The next test considers 11-layer graphene under biaxial compression. Figure \ref{fig:biax_11_3pc_orig_def} shows the out-of-plane deformations of the top layer of this system and cross-sections under $3\%$ strain. On inspection, the deformations appear to have similar morphology based on the top layer view and the presence of similar features and profiles in the cross-sections. This demonstrates the success of the ML-foliation model in reproducing the major deformation characteristics present in the foliation prediction. Comparing the 2D Fourier spectra in Figure \ref{fig:biax_11_3pc_fft_comp_img} provides further evidence of this.

\begin{figure}[htbp]
    \centering
    \includegraphics[width=0.35\linewidth]{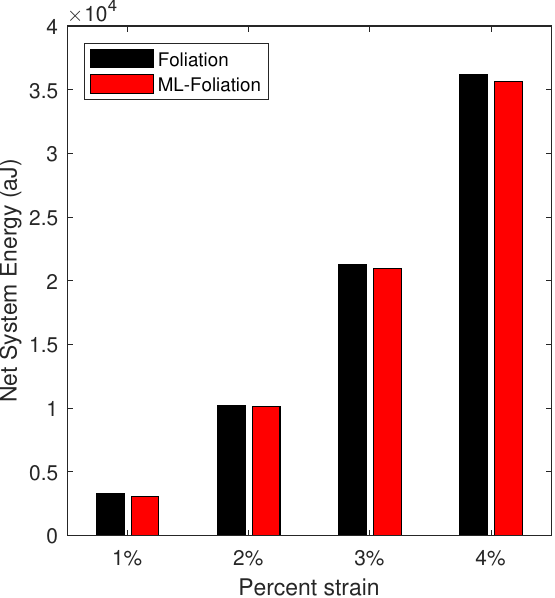}
    \includegraphics[width=0.353\linewidth]{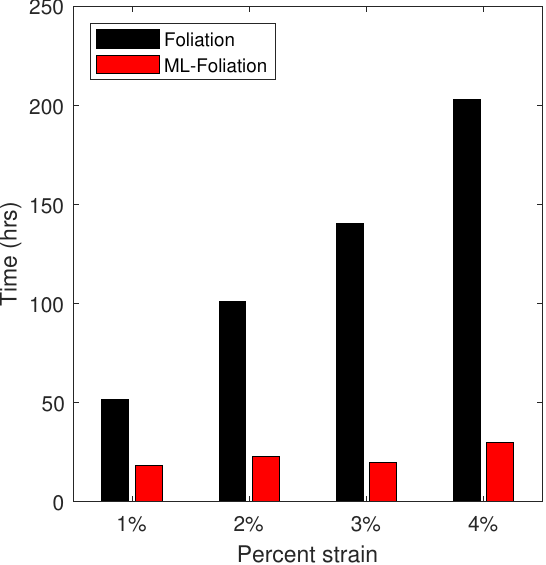}
    \caption{11-layer graphene under biaxial compression: Energy and time comparison for foliation model and ML-foliation model, denoted by black and red bars, respectively. }
    \label{fig:biax_11_ener_time_bars}
\end{figure}

Figure \ref{fig:biax_11_ener_time_bars} shows the energy and computation time for the predictions. As with 4-layer graphene, comparing the energy results indicates strong agreement between the foliation and ML-foliation predictions. Additionally, the ML-foliation model demonstrates an improvement in computational efficiency over the foliation model, though not as high as the uniaxial compression loading case. The proposed model achieves a peak computational speedup of $7.2\times$, with the lowest strain delivering a modest speedup of $2.8\times$. As with uniaxial compression, lower strains give a lesser speedup compared to higher strains due to shorter foliation model run times. However, the peak speedup is not necessarily observed at the highest strain. Systems under biaxial compression experience pure compression at the edges due to the boundary conditions. This leads to the formation of smaller wrinkles near the edges, which facilitate the transition from the flat, compressed graphene at the boundaries to larger folds at the center. The complexity of this phenomenon increases with increasing strain. The ML-foliation predictions may fail to accurately capture this complex deformation due to the small dataset and the approximations used in the model. This may explain the tendency of the ML-foliation model to converge more slowly at larger strains, such as $4\%$ (see Figures~\ref{fig:biax_4_ener_time_bars} and \ref{fig:biax_11_ener_time_bars}).

\begin{figure}[htbp]
    \centering
    \includegraphics[width=0.44\linewidth]{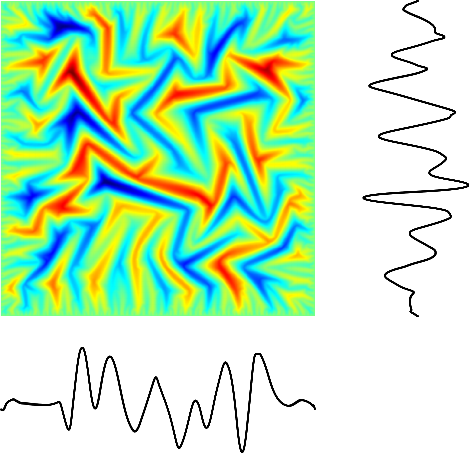}
    \includegraphics[width=0.545\linewidth]{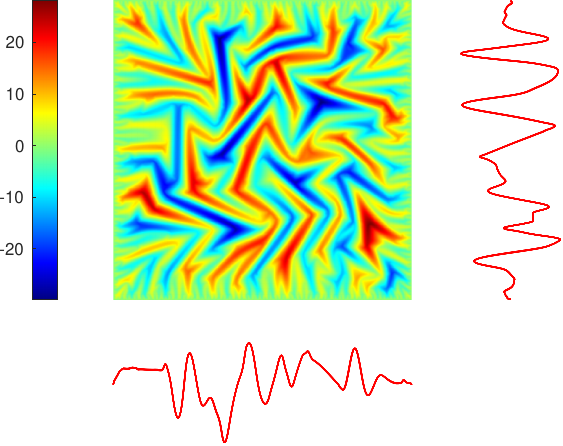}
    \caption{18-layer graphene under 3\% biaxial compression: out-of-plane deformation of the top layer predicted by the foliation model (left) and ML-foliation model (right), including central cross-sections, with the cross-sections to the right being in the $x_2$-$x_3$ plane, and the lower cross-sections being in the $x_1$-$x_3$ plane.}
    \label{fig:biax_18_def}
\end{figure}

\begin{figure}[htbp]
    \centering
    \includegraphics[width=0.3\linewidth]{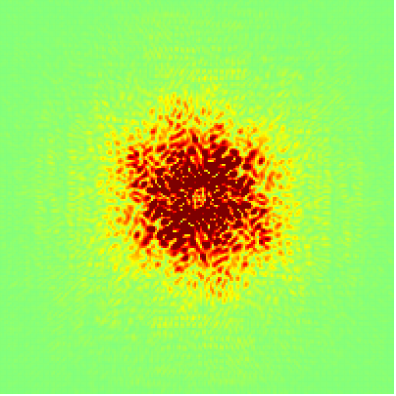}
    \includegraphics[width=0.41\linewidth]{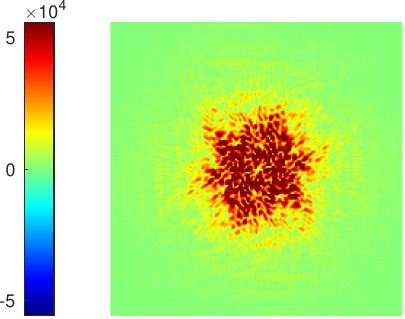}
    \caption{18-layer graphene under 3\% biaxial compression: Comparison of 2D Fourier amplitude spectra of the out-of-plane deformation obtained by the foliation model (left) and the ML-foliation model (right).}
\label{fig:biax_18_fft_comp_img}
\end{figure}

\begin{figure}[htbp]
    \centering
    \includegraphics[width=0.35\linewidth]{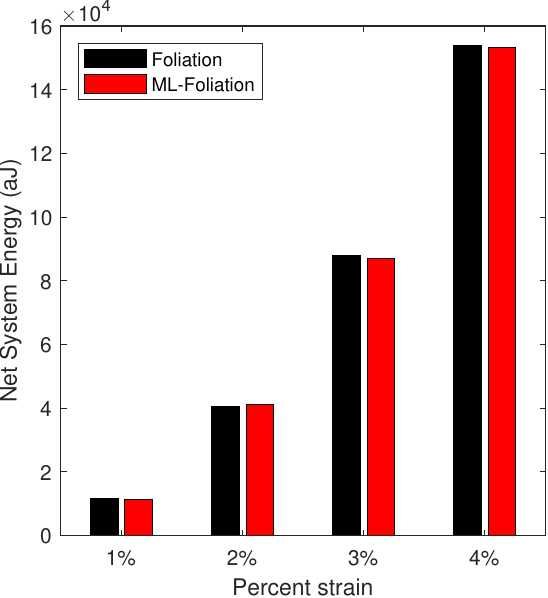}
    \includegraphics[width=0.357\linewidth]{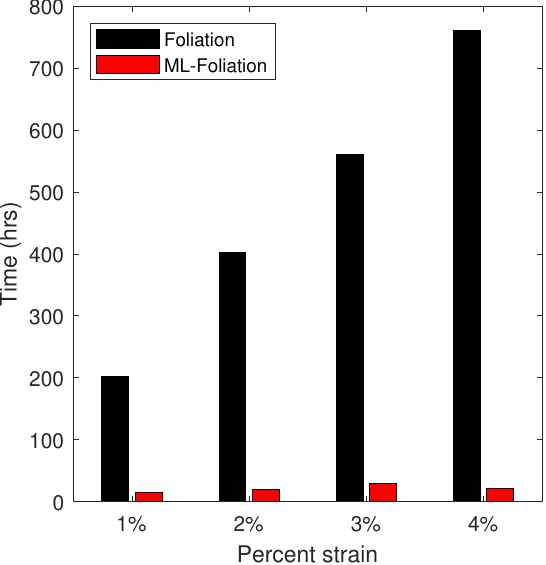}
    \caption{18-layer graphene under 3\% biaxial compression: Energy and time comparison for foliation model and ML-foliation model, denoted by black and red bars, respectively.}
    \label{fig:biax_18_ener_time_bars}
\end{figure}

The final test under biaxial compression considers 18-layer graphene. The out-of-plane deformation fields at $3\%$ strain are presented in Figure~\ref{fig:biax_18_def}, including the cross-sections at the center. The comparison of the deformation fields and cross-sections shows a strong match in morphology between the ML-foliation prediction and the foliation prediction. This is supported by the comparison of the Fourier spectra in Figure~\ref{fig:biax_18_fft_comp_img}. In addition, from the predicted energies and computation times shown in Figure~\ref{fig:biax_18_ener_time_bars}, the energy match is also quite strong while showing improved computational efficiency of the ML-foliation model, with speedups between $13.7\times$ and $36.4\times$, varying with strain.

\subsection{Biaxial Compression with Simple Shear}

Assessing the model against one final load case, biaxial compression combined with simple shear, serves to demonstrate the efficacy of the ML-foliation model on a diverse array of deformation patterns due to varied loading conditions on multilayer graphene systems. This combined loading produces deformation patterns with greater complexity than shear alone while remaining distinct enough from purely biaxial compression to warrant investigation. Like biaxial compression, this load case is computationally prohibitive to test beyond $4\%$ strain due to the complexity of the deformations and given the hardware constraints and maximum queue times on the shared high-performance computing infrastructure used in this work; thus, the biaxial compression with simple shear training data is also limited to $4\%$ strain.

\begin{figure}[htbp]
    \centering
    \includegraphics[width=0.32\linewidth]{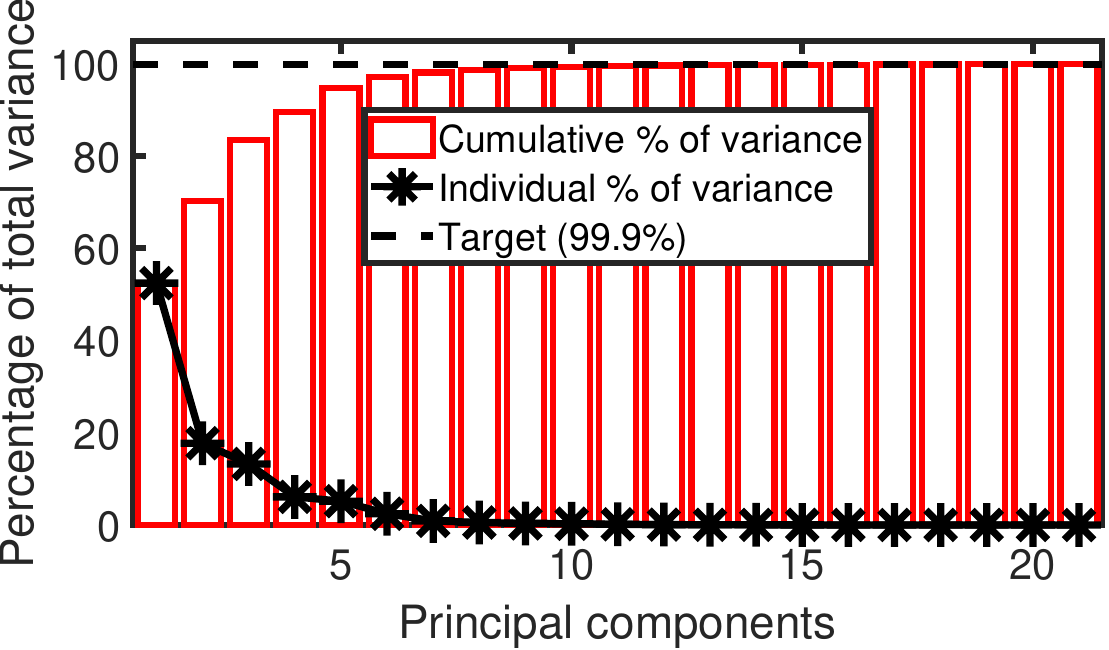}
    \includegraphics[width=0.32\linewidth]{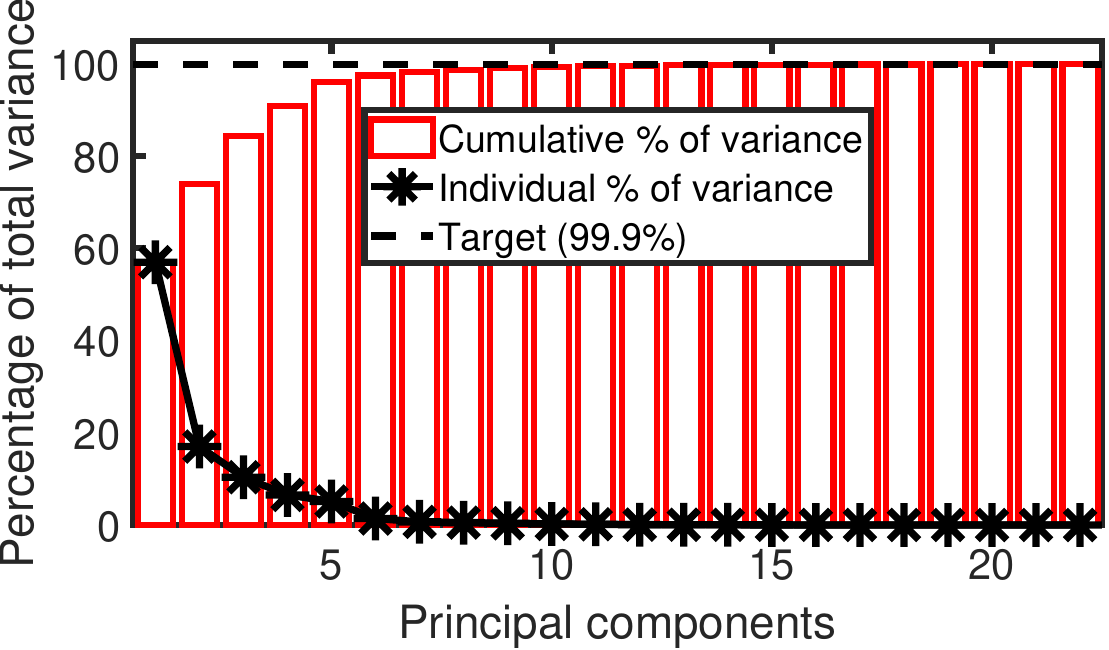}
    \includegraphics[width=0.32\linewidth]{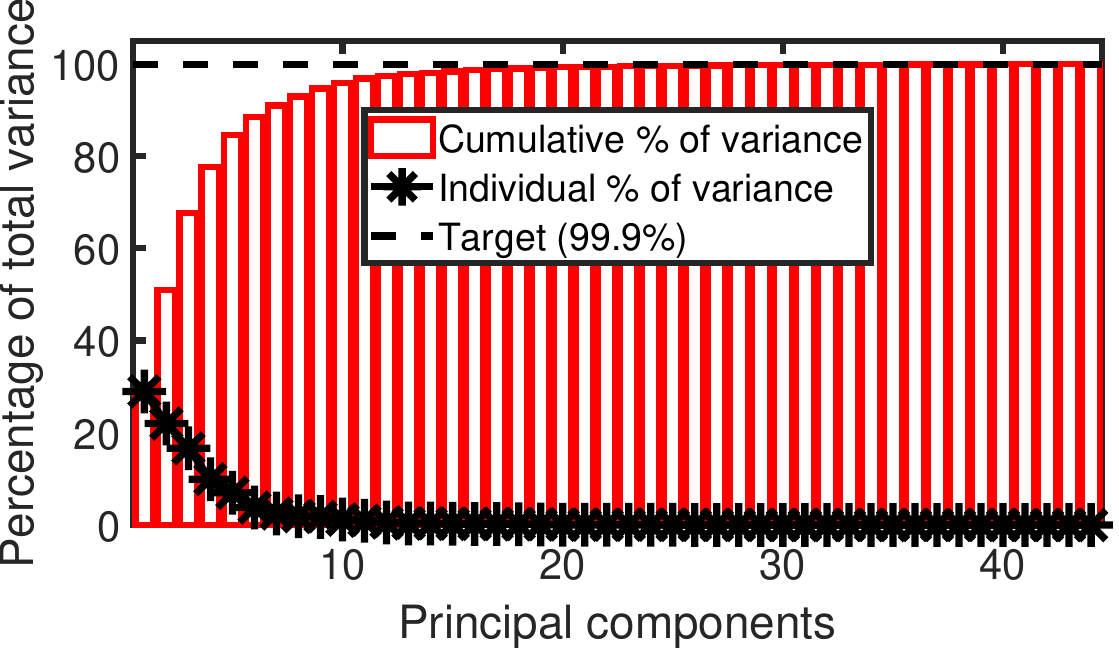}
    \caption{Scree and cumulative variance plots for FPCA of the biaxial compression with shear loading case. The horizontal axis represents the order of the principal components, while the vertical axis shows the cumulative and individual percentage of variance of each principal component, represented via red bars and black lines, respectively. The three plots from left to right correspond to the $x_1$, $x_2$, and $x_3$ deformation components. The total number of principal components in each plot is the number required to represent 99.9\% cumulative variance, as indicated by the black dashed line.}
    \label{fig:kle_variance_sshr_bi_gpn}
\end{figure}

FPCA is performed on the deformation data once again to obtain a new set of principal eigenfunctions. The scree plots are presented in Figure \ref{fig:kle_variance_sshr_bi_gpn}. This load case requires 21, 22, and 44 principal components to capture $99.9\%$ of the variance in the $x_1$, $x_2$, and $x_3$ direction deformation fields, respectively. This is quite similar to the biaxial results, differing very slightly in the number of required PCs, and the reduction in dimensionality remains equally high, roughly four orders of magnitude. Here, the first principal component in each direction captures between $29\%$ and $57\%$ of the total variance, with the higher percentages being in the $x_1$ and $x_2$ deformation components. This suggests that these components exhibit less complex deformation patterns than the $x_3$ component. Training a neural network on the PC scores from this load case also gives a Pearson correlation coefficient above 0.999.

\begin{figure}[htbp]
    \centering
    \includegraphics[width=0.44\linewidth]{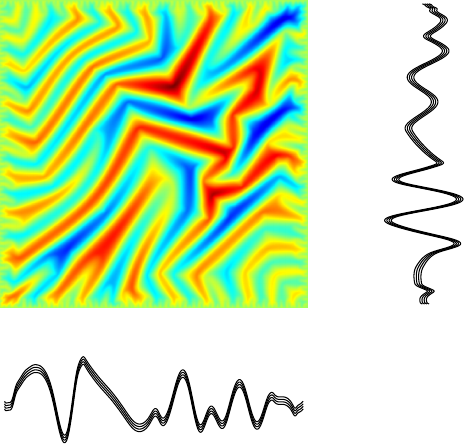}
    \includegraphics[width=0.53\linewidth]{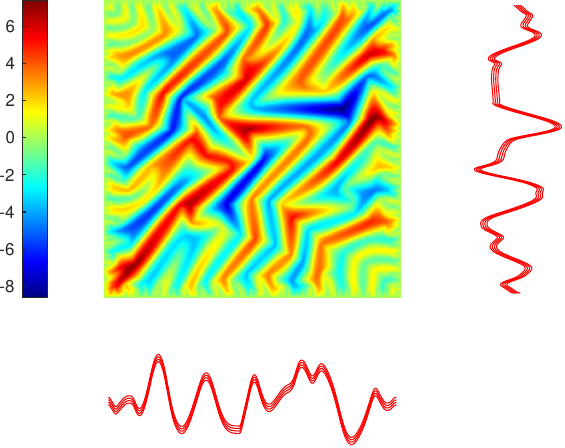}
    \caption{4-layer graphene under 3\% biaxial compression with shear: out-of-plane deformation of the top layer predicted by the foliation model (left) and ML-foliation model (right), including central cross-sections, with the cross-sections to the right being in the $x_2$-$x_3$ plane, and the lower cross-sections being in the $x_1$-$x_3$ plane.}
    \label{fig:sshr_bi_4_3pc_orig_def}
\end{figure}

\begin{figure}[htbp]
    \centering
    \includegraphics[width=0.3\linewidth]{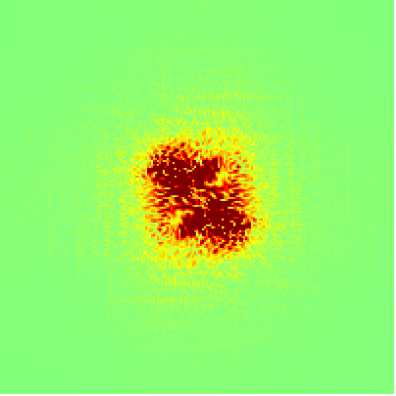}
    \includegraphics[width=0.425\linewidth]{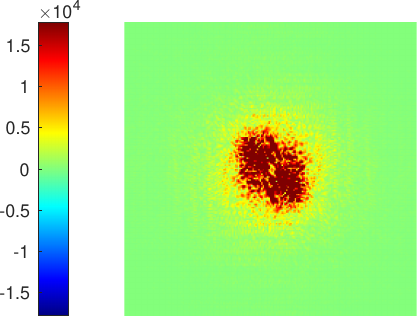}
    \caption{4-layer graphene under 3\% biaxial compression with shear: Comparison of 2D Fourier amplitude spectra of the out-of-plane deformation obtained by the foliation model (left) and the ML-foliation model (right).}
\label{fig:sshr_bi_4_3pc_fft_comp_img}
\end{figure}

The results for 4-layer graphene under biaxial compression with shear are as follows. Figure \ref{fig:sshr_bi_4_3pc_orig_def} shows the foliation and ML-foliation predictions of the out-of-plane deformation fields at the top layer of 4-layer graphene at 3\% strain with cross-section views. The morphologies of each are comparable, both by visual inspection and by comparing the profiles of the 2D Fourier spectra in Figure \ref{fig:sshr_bi_4_3pc_fft_comp_img}. The cross sections also show a match in the approximate wavelength of the deformations. Again, $3\%$ strain is presented because $4\%$ tends to produce poorer initial guesses for this load case.

\begin{figure}[htbp]
    \centering
    \includegraphics[width=0.35\linewidth]{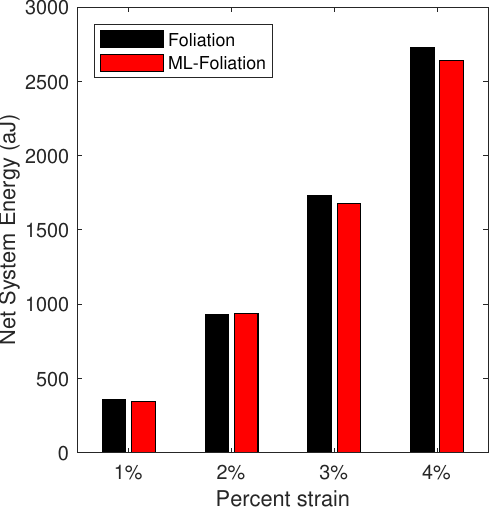}
    \includegraphics[width=0.35\linewidth]{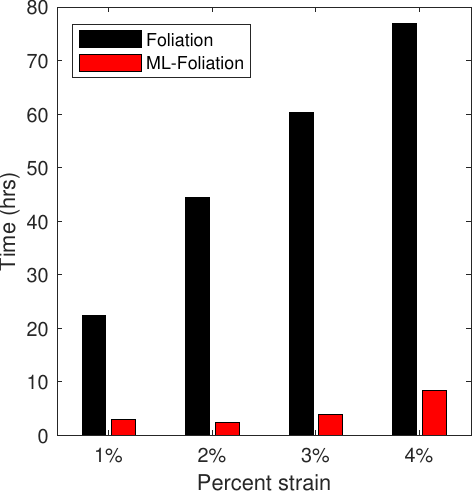}
    \caption{4-layer graphene under biaxial compression with shear: Energy and time comparison for foliation model and ML-foliation model, denoted by black and red bars, respectively.}
    \label{fig:sshr_bi_4_ener_time_bars}
\end{figure}

The energy and computational cost metrics for 4-layer graphene under biaxial compression with shear are presented in Figure \ref{fig:sshr_bi_4_ener_time_bars}. As with biaxial compression, the comparison of the two approaches indicates a strong energy match. The ML-foliation model offers computational speedups from $7.8\times$ up to $19.0\times$, varying across the tested strain levels.

\begin{figure}[htbp]
    \centering
    \includegraphics[width=0.44\linewidth]{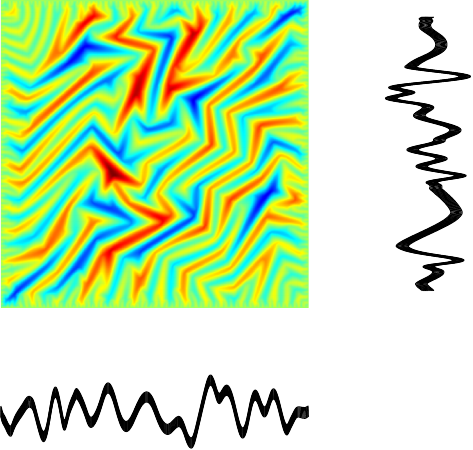}
    \includegraphics[width=0.554\linewidth]{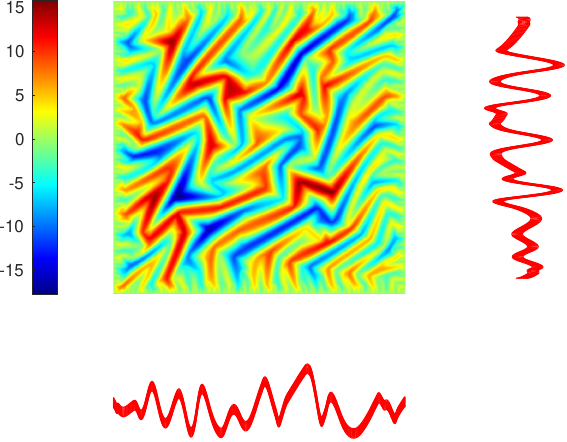}
    \caption{11-layer graphene under 3\% biaxial compression with shear: The foliation model (left) and ML-foliation model (right) results for the out-of-plane deformation at the top layer with the cross sections at the central planes, the rightward cross sections being in the $x_2$-$x_3$ plane, and the lower cross sections being in the $x_1$-$x_3$ plane. }
    \label{fig:sshr_bi_11_3pc_orig_def}
\end{figure}

\begin{figure}[htbp]
    \centering
    \includegraphics[width=0.3\linewidth]{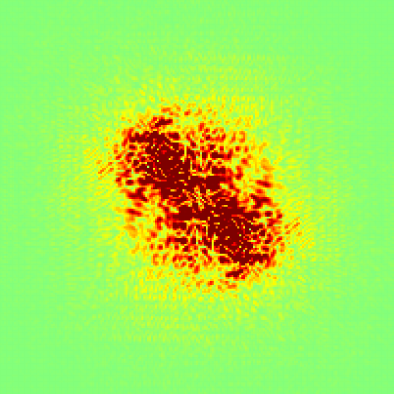}
    \includegraphics[width=0.41\linewidth]{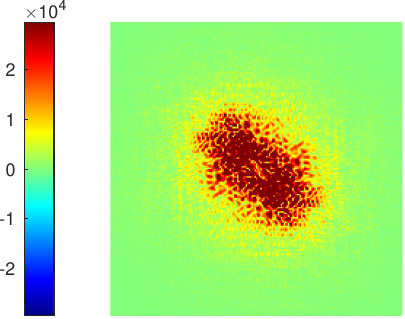}
    \caption{11-layer graphene under 3\% biaxial compression with shear: Comparison of 2D Fourier amplitude spectra of the out-of-plane deformation obtained by the foliation model (left) and the ML-foliation model (right).}
\label{fig:sshr_bi_11_3pc_fft_comp_img}
\end{figure}

The next set of results for this load case assesses the performance of the proposed model on 11-layer graphene under biaxial compression with shear. Figure \ref{fig:sshr_bi_11_3pc_orig_def} shows the out-of-plane deformations at the top layer of this system at 3\% strain with cross-sections. Once again, the deformations exhibit similar morphology, sharing several features like the variation in wavelength and double peaks seen in the cross-sections. The similarity of the general profiles of the Fourier spectra in Figure \ref{fig:sshr_bi_11_3pc_fft_comp_img} provides further evidence of the morphology match.

\begin{figure}[htbp]
    \centering
    \includegraphics[width=0.35\linewidth]{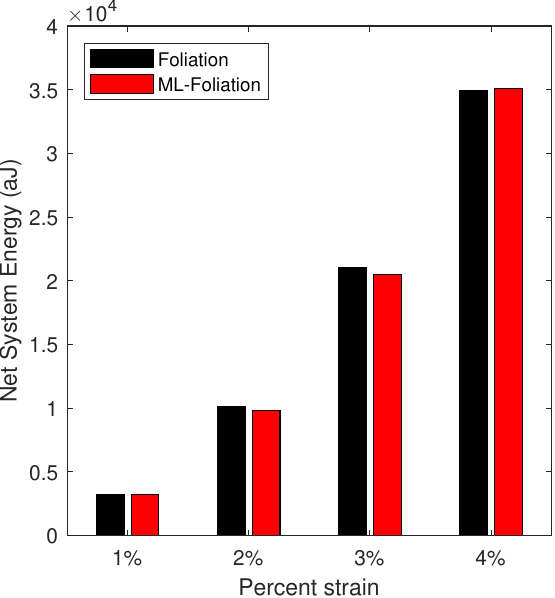}
    \includegraphics[width=0.353\linewidth]{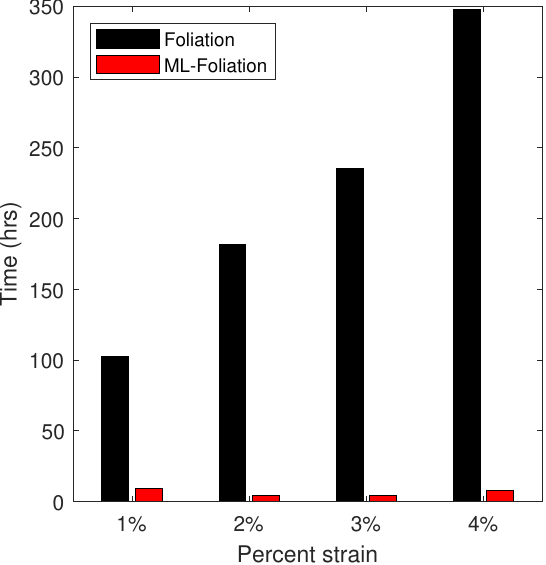}
    \caption{11-layer graphene under biaxial compression with shear: Energy and time comparison for foliation model and ML-foliation model, denoted by black and red bars, respectively.}
    \label{fig:sshr_bi_11_ener_time_bars}
\end{figure}

Figure \ref{fig:sshr_bi_11_ener_time_bars} compares the energy and computational costs of the two approaches for 11-layer graphene under biaxial compression with shear between the foliation and ML-foliation predictions. These energy results are also highly consistent, as was the case for 4-layer. The ML-foliation model once again demonstrates higher computational efficiency than the foliation model, with a speedup ranging from $11.0 \times$ to $54.7\times$ across different strains. This is the highest speedup of all the test cases and can be attributed to the elevated computation time for this load case and the ML-foliation model's high computational efficiency in the minimization.

\begin{figure}[htbp]
    \centering
    \includegraphics[width=0.44\linewidth]{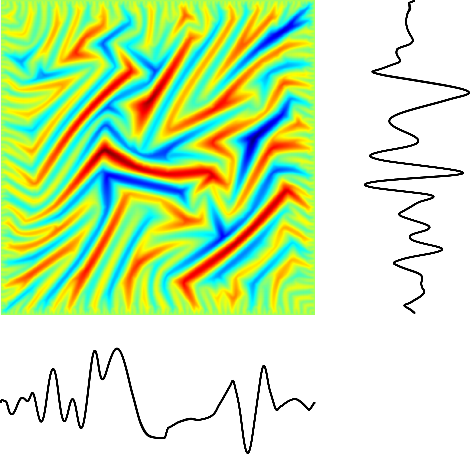}
    \includegraphics[width=0.545\linewidth]{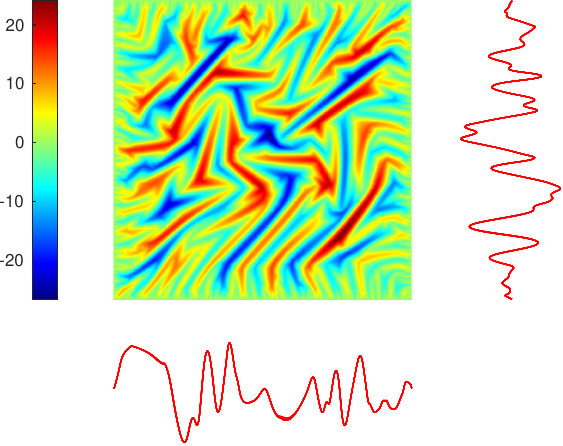}
    \caption{18-layer graphene under 3\% biaxial compression with shear: out-of-plane deformation of the top layer predicted by the foliation model (left) and ML-foliation model (right), including central cross-sections, with the cross-sections to the right being in the $x_2$-$x_3$ plane, and the lower cross-sections being in the $x_1$-$x_3$ plane.}
    \label{fig:sshr_bi_18_def}
\end{figure}

\begin{figure}[htbp]
    \centering
    \includegraphics[width=0.3\linewidth]{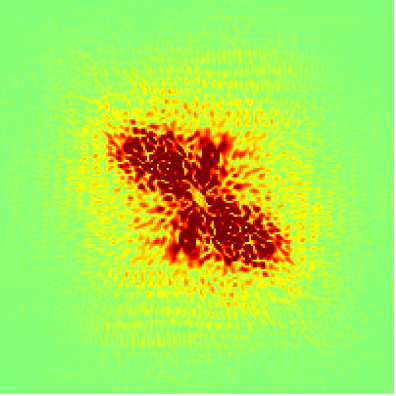}
    \includegraphics[width=0.41\linewidth]{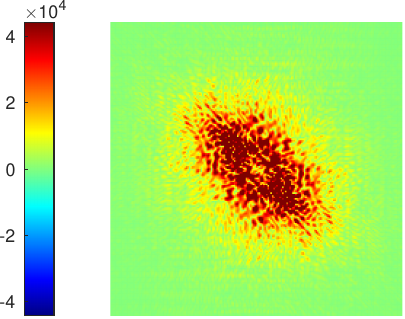}
    \caption{18-layer graphene under 3\% biaxial compression with shear: Comparison of 2D Fourier amplitude spectra of the out-of-plane deformation obtained by the foliation model (left) and the ML-foliation model (right).}
\label{fig:sshr_bi_18_fft_comp_img}
\end{figure}

\begin{figure}[htbp]
    \centering
    \includegraphics[width=0.35\linewidth]{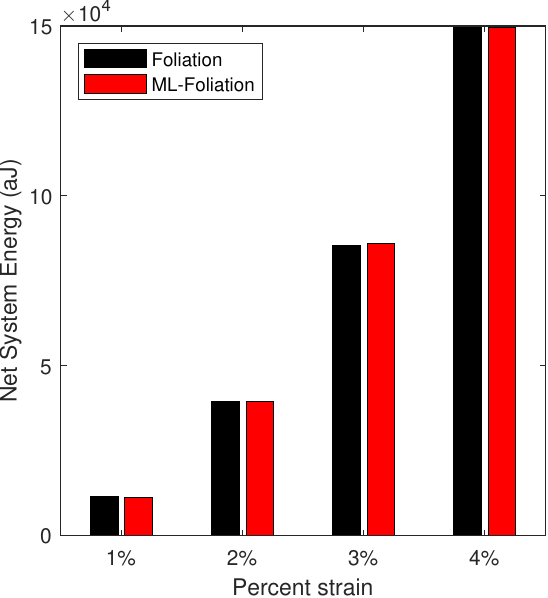}
    \includegraphics[width=0.358\linewidth]{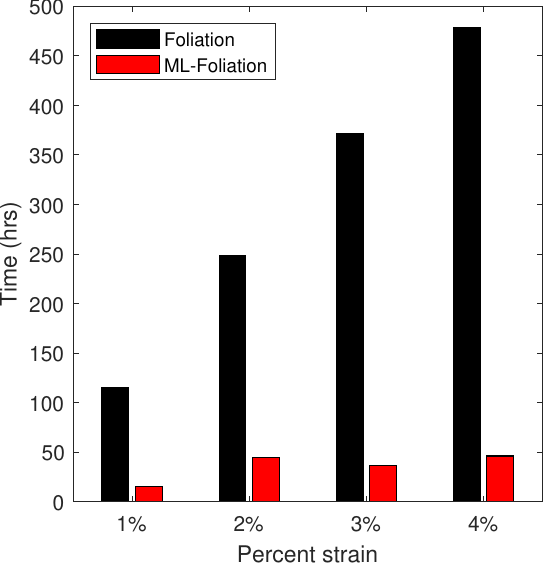}
    \caption{18-layer graphene under biaxial compression with shear: Energy and time comparison for foliation model and ML-foliation model, denoted by black and red bars, respectively.}
    \label{fig:sshr_bi_18_ener_time_bars}
\end{figure}

The last test for this load case considers 18-layer graphene. The out-of-plane deformation fields and central cross-sections at $3\%$ strain are presented in Figure~\ref{fig:sshr_bi_18_def}. Once again, the morphology is very similar, with the cross-sections sharing several unique features. Comparing the Fourier spectra in Figure~\ref{fig:sshr_bi_18_fft_comp_img} shows reduced similarity, as the 2D Fourier spectrum of the foliation prediction begins to lose uniformity in its profile while the spectrum of the ML-foliation prediction maintains a profile similar to smaller system sizes. This suggests it is interpolating the general features but cannot extract the more precise details for this system size. Still, the predicted energies and computation times shown in Figure~\ref{fig:sshr_bi_18_ener_time_bars} support the validity of the ML-foliation results. The energy values are incredibly similar, and the ML-foliation model achieves a computational speedup between $7.5\times$ and $10.4\times$, varying with strain. 

\section{Explainability}
Machine learning models are often criticized for their “black-box” nature, which can obscure the decision-making process and reduce confidence in their predictions \cite{rudin2019stop}. Here, we briefly discuss the interpretability aspect of the proposed ML framework. The ML model development consists of two steps. First, the eigenfunctions of the data covariance operator are determined to represent the data in terms of the principal components. This is followed by learning the mapping between the system inputs of a deformed configuration and the corresponding principal components. Each deformation field is consequently expressed as a linear combination of these eigenfunctions. This representation facilitates interpretability by identifying dominant deformation modes (eigenfunctions) and quantifying their relative contributions through the learned coefficients, providing a physically meaningful basis for identifying the major wrinkling patterns. Therefore, the internal mechanics and prediction-making steps of the ML model can be easily understood by the user, since the predictions are made in terms of eigenfunctions that are visually interpretable. 

\begin{figure}[htbp]
    \centering
    \includegraphics[width=0.24\linewidth]{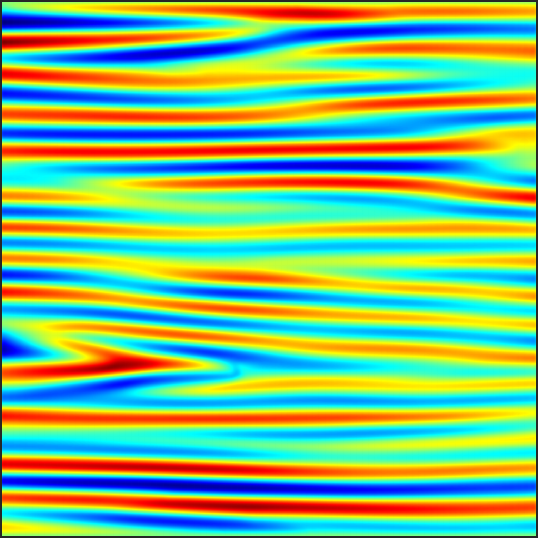}
    \includegraphics[width=0.24\linewidth]{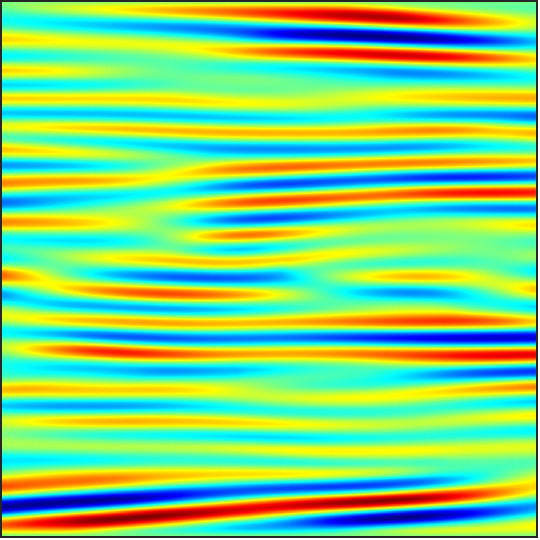}
    \includegraphics[width=0.24\linewidth]{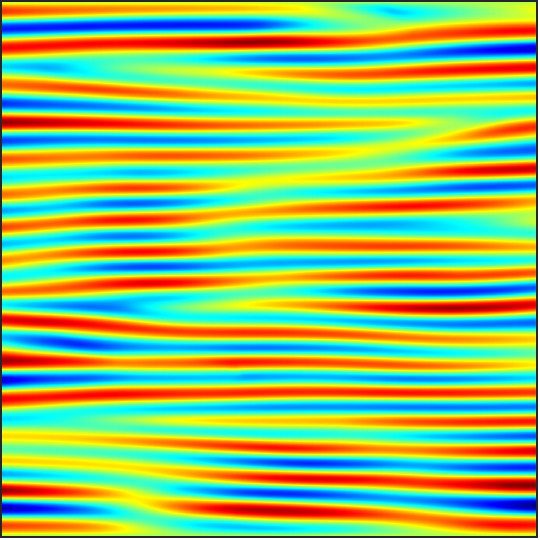}
    \includegraphics[width=0.24\linewidth]{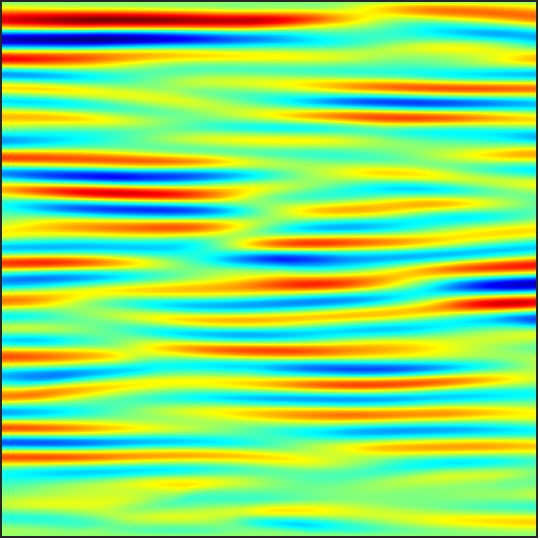}
    \includegraphics[width=0.24\linewidth]{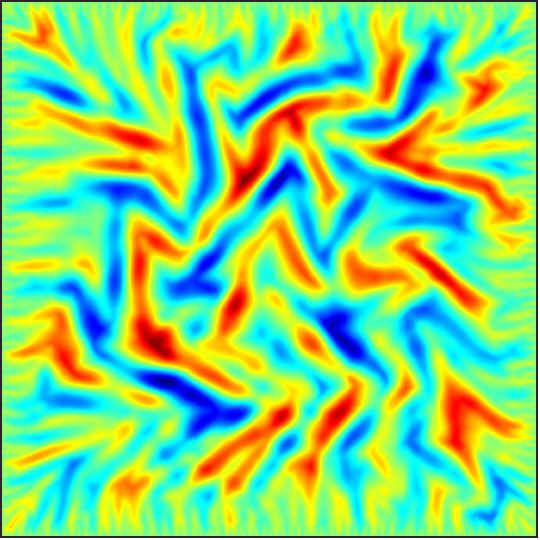}
    \includegraphics[width=0.24\linewidth]{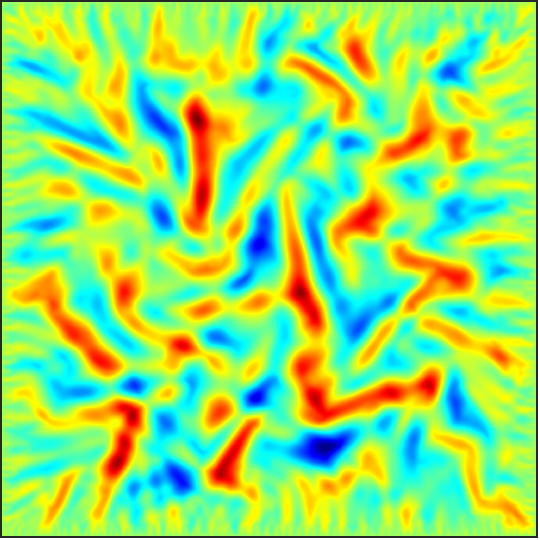}
    \includegraphics[width=0.24\linewidth]{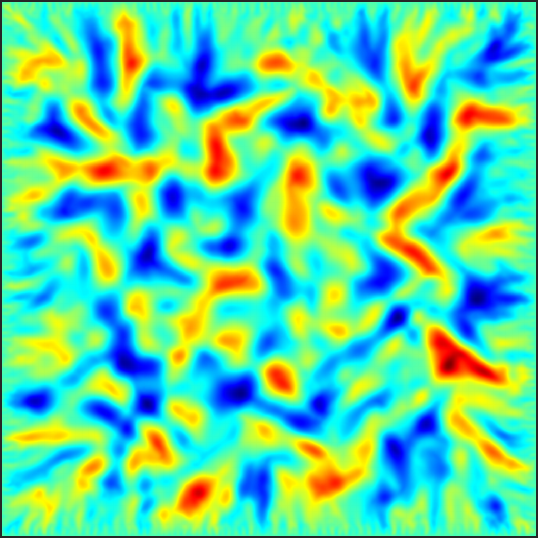}
    \includegraphics[width=0.24\linewidth]{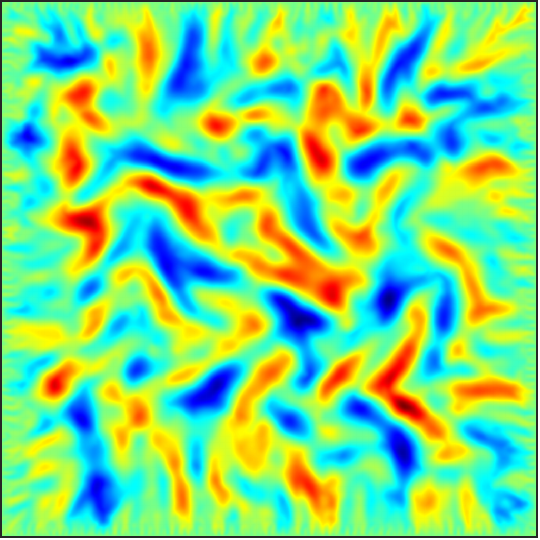}
    \includegraphics[width=0.24\linewidth]{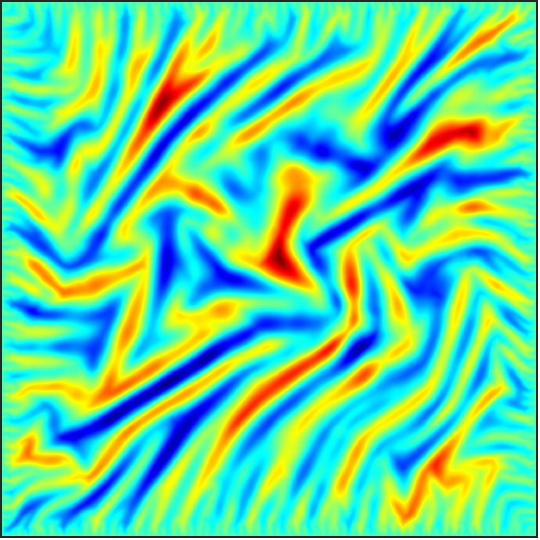}
    \includegraphics[width=0.24\linewidth]{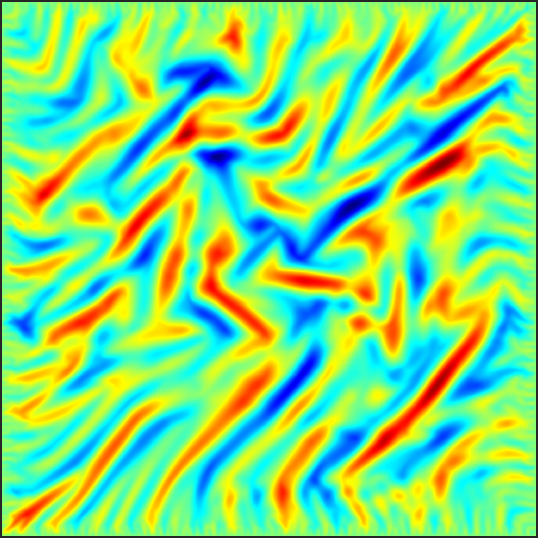}
    \includegraphics[width=0.24\linewidth]{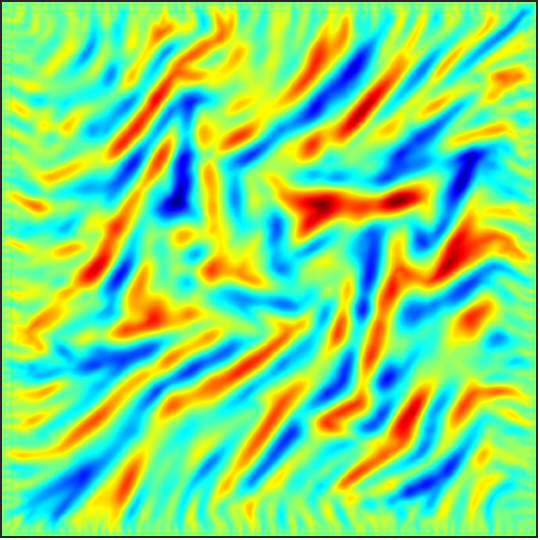}
    \includegraphics[width=0.24\linewidth]{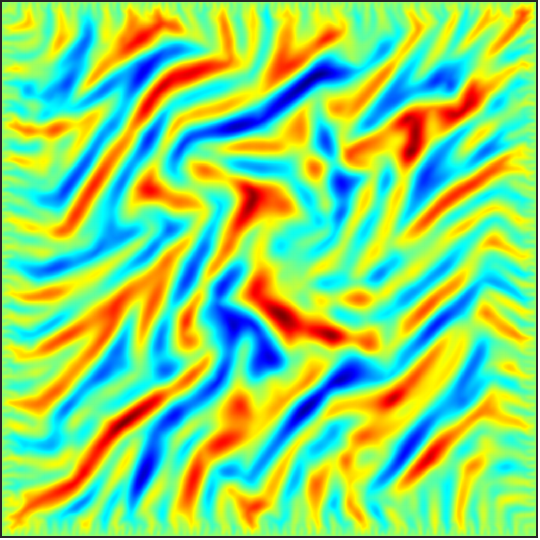}
    \caption{The first four principal components of displacement in the $x_3$ direction for each load case. Principal components are shown in ascending order from left to right, starting with the first principal component on the far left. The rows from top to bottom correspond to uniaxial compression, biaxial compression, and biaxial compression with shear, respectively.}
\label{fig:all_pcs}
\end{figure}

A few eigenfunctions for the uniaxial compression case are presented here to visualize key deformation patterns in the top row of Figure~\ref{fig:all_pcs}. These first four eigenfunctions capture $78.6\%$ of the variance in the $x_3$ deformation data. The first eigenfunction captures both wrinkle structures characterized by straight folds and their bifurcation points, consistent with the deformations shown earlier (Figures~\ref{fig:uniax_4_def}, \ref{fig:uniax_11_def}). The second,  third, and fourth eigenfunctions similarly represent dominant wrinkle modes, with enhanced emphasis on their termination regions, where the folds gradually smooth out or bifurcate at various spatial locations. As a result of this decomposition, realistic deformations can be well approximated by linear combinations of a small number of these modes.

The biaxial compression case also reveals principal components that admit a clear physical interpretation within a mode-decomposition framework (see middle row of Figure~\ref{fig:all_pcs}). In this case, the first four eigenfunctions capture $79.0\%$ of the variance in the $x_3$ deformation data. The first eigenfunction closely aligns with the dominant biaxial deformation pattern (as shown in Figures~\ref{fig:biax_4_3pc_orig_def}, \ref{fig:biax_11_3pc_orig_def}), capturing the primary wrinkle morphology containing both the folds and junctions of the folds. The second, third, and fourth eigenfunction represent deformation modes characterized by wrinkled structures with more pronounced details of junctions of wrinkles rather than straight folds. They also isolate junction-dominated features of varying shape and amplitude at different locations and orientations, indicating that these intersection regions constitute distinct deformation modes. They also exhibit a range of positive and negative values of displacements. The first and second eigenfunctions show more small wrinkles near the edges than the third and fourth. The eigenfunctions for the biaxial compression case exhibit more interrupted folds than those for the uniaxial compression case. This is attributed to the greater number of fold intersections in the deformation resulting from the biaxial compression.

The last load case, biaxial compression with simple shear, also demonstrates that the principal components provide a realistic analog to a mode-decomposition framework (see bottom row of Figure~\ref{fig:all_pcs}). Here, the first four eigenfunctions capture $77.6\%$ of the variance in the $x_3$ deformation data. The first eigenfunction possesses many of the dominant features present in the shear with biaxial deformation patterns (as shown in Figures~\ref{fig:sshr_bi_4_3pc_orig_def}, \ref{fig:sshr_bi_11_3pc_orig_def}), including simple wrinkles at the edges and corners with more complex features in the center. The second to fourth eigenfunctions capture wrinkles largely parallel to the primary axis of wrinkling, along with an assortment of splits and junctions between the wrinkles that are more prominent near the center. All four eigenfunctions also exhibit varying degrees of emphasis on wrinkles near the edges. 

\section{Discussion and Conclusion} 
In this work, we developed a machine-learning framework to accelerate atomistic-continuum simulations of multilayer crystalline membranes. The framework applies functional principal component analysis (FPCA) directly to the 2D deformed surfaces of graphene layers, reducing the dimensionality of the deformation fields by several orders of magnitude. A feedforward neural network is then used to learn the corresponding low-dimensional FPCA representation from the system and loading parameters. The ML-predicted deformation serves as an informed initial guess for the foliation model, which performs energy minimization to obtain an equilibrium atomistic-continuum solution. The resulting ML-foliation solutions closely reproduce the overall wrinkle morphology and equilibrium energies obtained using the conventional foliation model. The framework is demonstrated for multilayer graphene (with the largest system containing about three billion atoms) subjected to uniaxial compression, biaxial compression, and combined biaxial compression and simple shear.

The proposed framework offers several advantages. First, it extends FPCA-based dimensionality reduction from collections of 1D curves to 2D deformation surfaces, thereby avoiding the spurious inter-curve discontinuities that can arise in 1D curve-based FPCA representations. Second, by constructing the functional representation using 2D basis functions that exactly satisfy the prescribed displacement boundary conditions, the corresponding truncated FPCA expansions also satisfy these constraints exactly. Third, the ML-assisted initialization substantially reduces the computational cost of the subsequent atomistic-continuum minimization. For computationally demanding large-strain cases, corresponding to approximately \(3\%-5\%\) applied strain, the ML-foliation model achieves speedups of at least \(6.8\times\). Furthermore, these predictions are equilibrium configurations with wrinkle morphologies comparable to those obtained via the foliation model and with energies that are similar to, and in some cases lower than, those obtained by the foliation model alone. Fourth, the FPCA representation provides an interpretable reduced-order description of the deformation fields by enabling visualization of the dominant deformation modes and quantification of their relative contributions.

An important feature of the proposed approach is that the machine-learning model is not required to predict a unique equilibrium wrinkle configuration. Because multilayer graphene has very low bending stiffness, the resulting wrinkle patterns are highly sensitive to small perturbations in the initial configuration, and repeated simulations of the same system may converge to different but energetically comparable equilibrium states. Consequently, the role of the ML model is to predict a deformation field that captures the dominant structural characteristics of the solution and provides a physically meaningful initialization for the atomistic-continuum solver. The subsequent foliation-model minimization then restores mechanical equilibrium. This hybrid formulation (ML-foliation) therefore adheres to the underlying physics-based energy model while using machine learning to accelerate convergence.

The proposed framework demonstrates strong predictive capabilities and substantial computational acceleration, even with the relatively limited size of the dataset. The dataset size is constrained by the high computational cost associated with generating high-fidelity atomistic-continuum data. These results highlight the data efficiency of the FPCA-based reduced-order representation and its ability to capture the dominant deformation characteristics across unseen systems. Further improvements in morphological agreement and broader generalization can be expected as larger and more diverse training datasets become available. Future work will also focus on developing unified learning models capable of representing multiple loading conditions within a single framework.

Although the present study focuses on multilayer graphene under three representative loading conditions, the proposed methodology is not restricted to any particular material or loading condition. The framework is potentially applicable to other multilayer two-dimensional crystalline materials and, more broadly, to computational mechanics problems in which high-dimensional solution fields exhibit coherent low-dimensional structure on two-dimensional manifolds. Such extensions could provide a general strategy for accelerating expensive physics-based simulations while retaining the accuracy and physical consistency of the underlying numerical model.

\section{Acknowledgments}
This research was primarily supported by the National Science Foundation under Grant No. 2442313, which provided support for all aspects of the work except the neural-network component. The U.S. Department of Energy, Office of Science, Award No. DE-SC0023432 supported the work on neural-network model development and its applications (AK, SP, and SG). Computational resources were provided by Bridges-2 at Pittsburgh Supercomputing Center through allocation CIS260992 from the Advanced Cyberinfrastructure Coordination Ecosystem: Services \& Support (ACCESS) program, which is supported by U.S. National Science Foundation grants \#2138259, \#2138286, \#2138307, \#2137603, and \#2138296. Additional computational resources were provided by Superior and MRI computing clusters at Michigan Technological University.

\section*{Competing Interests}
\noindent The authors declare no competing financial or non-financial interests.

\section*{Data Availability}\label{subsec:Data_availability}
\noindent Raw data were generated at Bridges-2 at Pittsburgh Supercomputing Center and Superior, a high-performance computing facility at Michigan Technological University. Derived data supporting the findings of this study are available from the corresponding author upon request.

\section*{Code Availability}\label{subsec:Code_availability}
Codes for FPCA supporting the findings of this study are available from the corresponding author upon reasonable request.

\section*{Declaration of generative AI and AI-assisted technologies in the manuscript preparation process}
During the preparation of this work, the author(s) used GPT-5 (OpenAI) for language editing and proofreading. The author(s) reviewed and edited the output as needed and take full responsibility for the content of the published article.

\appendix
\section{Neural network architecture details}
The following table summarizes the architecture of the neural network employed in this work, including the number of layers and neurons in each layer, among other details.

\begin{table}[htbp]
\centering
\begin{tabular}{@{}llll@{}}
\toprule
Layer & Type & Output size & Activation \\
\midrule
Input     & --     & 3                              & --            \\
Hidden 1  & Linear & 500                            & ReLU          \\
Hidden 2  & Linear & 1000                           & ReLU          \\
Hidden 3  & Linear & 1000                           & ReLU          \\
Hidden 4  & Linear & 1000                           & ReLU          \\
Hidden 5  & Linear & 2000                           & ReLU          \\
Hidden 6  & Linear & 500                            & ReLU          \\
Output    & Linear & $K$ (no. of fPCA components)   & Linear (none) \\
\bottomrule
\end{tabular}
\caption{Deep neural network architecture.}
\label{tab:nn_architecture}
\end{table}

\bibliographystyle{elsarticle-num} 
\bibliography{main.bib}

\end{document}